\documentclass[
    prx,
    twocolumn,
    superscriptaddress,
    nofootinbib
]{revtex4-2}

\usepackage[english]{babel}
\usepackage[dvipsnames]{xcolor}
\usepackage{cancel}
\usepackage{hyperref}
\usepackage{amsmath}
\usepackage[normalem]{ulem}
\usepackage{placeins} 

\hypersetup{
  breaklinks=true,
  colorlinks=true,
  allcolors=BlueViolet,
}
\usepackage{booktabs}
\usepackage{dsfont}
\usepackage{soul}

\usepackage{graphicx}

\usepackage{physics,braket,bm}

\usepackage{amssymb}

\usepackage{standalone}

\usepackage[inline]{enumitem}
\setlist[enumerate]{leftmargin=*}

\newcommand{\UWM}{Department of Physics, University of Wisconsin-Madison, Madison, WI 53562, USA}
\newcommand{\EPFL}{Institute of Physics, Ecole Polytechnique Federale de Lausanne, 1015 Lausanne, Switzerland}
\newcommand{\UMD}{Department of Physics, Joint Quantum Institute, and Quantum Materials Center, University of Maryland, College Park, MD 20742, USA}

\begin{document}

\title{Proposal for erasure conversion in integer fluxonium qubits}
\author{Jiakai Wang}
\affiliation{\UWM}
\author{Raymond A. Mencia}
\affiliation{\EPFL}
\affiliation{\UMD}
\author{Vladimir E. Manucharyan}
\affiliation{\EPFL}
\affiliation{\UMD}
\author{Maxim G. Vavilov}
\affiliation{\UWM}
\date{\today}

\begin{abstract}
We propose an erasure conversion scheme on the $\ket{e}-\ket{f}$ and $\ket{g}-\ket{f}$ qubits in integer fluxonium qubits (IFQs), which are both first-order insensitive to $1/f$ flux noise. The $\ket{e}-\ket{f}$ transition is identical to that of a usual fluxonium qubit and hence is expected to have excellent coherence time, while the $\ket{g}-\ket{f}$ transition is additionally protected from the energy relaxation by the parity symmetry. The dominant error in both qubits arises due to the energy relaxation: from $\ket{e}$ to $\ket{g}$ in the e–f qubit and from $\ket{f}$ to $\ket{e}$ in the g–f qubit. Such errors can be treated as erasure events, and their efficient detection improves the performance of quantum error-correcting codes. We consider a protocol for such erasure conversion based on the dispersive readout. Our main finding is that, with proper circuit parameter choice, carefully designed gate sets, and the integration of erasure conversion, IFQs promise high effective coherence times.
\end{abstract}

\maketitle

\section{Introduction}\label{sec:intro}

\begin{figure*}[t]
    \centering
    \includegraphics[width=1\linewidth]{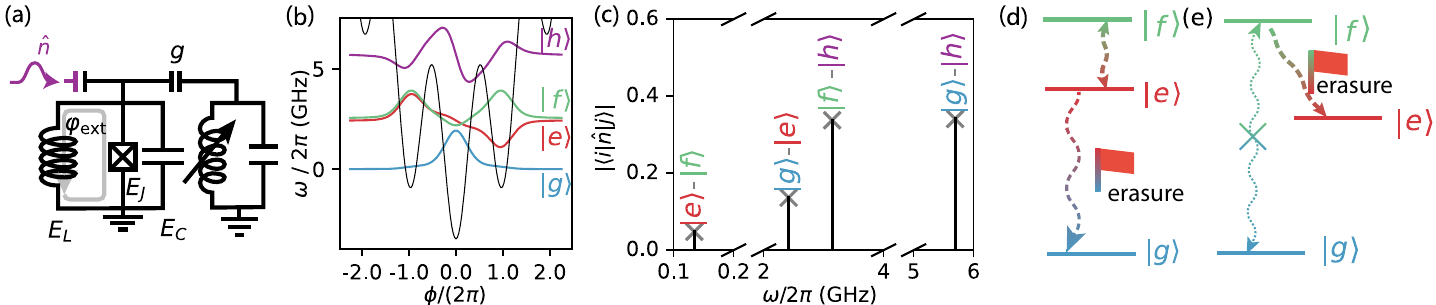}
    \caption{\textbf{(a)} Circuit diagram of a fluxonium capacitively coupled to a frequency-tunable resonator. When the resonator is populated with photons from the microwave drive, the output photon from the resonator's decay carries away information about the fluxonium state. \textbf{(b)} Potential landscape and wavefunctions of fluxonium at integer flux bias. The depicted wavefunctions correspond to the example g–f qubit considered in this work. The degeneracy of the two states localized in $m=\pm 1$ potential wells is lifted by consecutive single-well tunneling of amplitude $\epsilon_1$. \textbf{(c)} The absolute value of the charge matrix elements within the lowest four levels of the example g-f qubit in Table \ref{tab:qubit_parameters}. The relatively small $\ket{e}-\ket{f}$ element ensures erasure error do not overwhelm the erasure conversion-correction protocol, the relatively large $\ket{g}-\ket{h}$, $\ket{f}-\ket{h}$ transition elements facilitate Raman gate operation via the intermediate state $\ket{h}$. For the e-f qubit, we want to use an even lower $\ket{e}-\ket{f}$ frequency to enhance qubit lifetime, therefore the $\ket{e}-\ket{f}$ charge matrix element will be significantly smaller. \textbf{(d)} In the e-f qubit, the computational level frequency lies below the environmental temperature, while the leakage channel $\ket{g}$-$\ket{e}$ frequency is above it. The primary computational errors are the decay of $\ket{f}$ to $\ket{e}$ and the heating of $\ket{e}$ to $\ket{f}$. Leakage occurs when $\ket{e}$ decays to $\ket{g}$. Symmetry forbids transitions between $\ket{f}$ and $\ket{g}$. \textbf{(e)}  For the g-f qubit, direct transitions between computational states are forbidden. The primary error in the computational subspace arises from dephasing induced by frequency curvature with respect to external flux. Leakage occurs when $\ket{f}$ decays to $\ket{e}$.}
    \label{fig:fig1}
\end{figure*}

Leveraging the noise structure can effectively reduce the cost of quantum error correction (QEC) in fault-tolerant quantum computing \cite{original_xzzx,puri2020bias_cat,cong2022hardware}. Specifically, erasure errors are easily correctable because their locations are known \cite{gottesman1997stabilizer, grassl1997codes}. For a fixed total error budget, a high ratio of erasure errors leads to elevated code thresholds and higher-order suppression of logical error rates \cite{1997erasure,2009erasure,Yue2022erasure,sahay2023biasederasure,kubica2023erasure,ion2023erasure}. In gate-model quantum computers, erasure errors are often realized by detecting leakage and converting leakage populations back to the computational subspace. Even when this conversion of leakage to erasure is not spatially or temporally perfect, erasure errors still perform better than traditional non-heralded errors \cite{kubica_Aug24_theory,spatial_temporal_imperfection_of_erasure_checks}. Thus, erasure qubits are promising candidates for achieving fault-tolerant quantum computing. 

Erasure qubits have been proposed and experimentally demonstrated based on natural atoms \cite{Yue2022erasure,ion2023erasure,sahay2023biasederasure,ma2023high} and composite superconducting qubits \cite{levine2024dualrail,chou2024dualrail, kou2024erasurefluxonium,levine2024dualrail,koottandavida2023dualrail,kumar2024protomon}; both with error rates near or below QEC thresholds. The implementations of erasure qubits based on superconducting artificial atoms include the dual-rail transmon \cite{kubica2023erasure,kubica2023heralding} and the 
Floquet fluxonium molecule \cite{kou2024erasurefluxonium}, both having coherent computational subspaces but an erasure lifetime about two orders of magnitude shorter than computational subspace lifetime. The dual-rail transmon qubit exhibits a low bit-flip error rate within the computational subspace due to its first-order forbidden bit-flip process, while the decoherence-free subspace mitigates phase-flip errors. The Floquet fluxonium molecule reduces bit-flip errors through non-overlapping wavefunctions and suppresses phase-flip errors by Floquet engineering. Because these composite superconducting erasure qubits have great potential for hardware-efficient fault-tolerant quantum computing, it is timely to answer the question of whether erasure qubits can be engineered based on the bare energy levels of a single-mode superconducting artificial atom, such that the complexity of erasure qubits can be reduced. 
 
In this paper, we answer this question by analyzing two different computational state configurations of fluxonium. Fluxonium is a potential contender as a building block for next-generation quantum processors due to its large anharmonicity, which alleviates frequency crowding \cite{roadmap}. Additionally, its $\varphi_\text{ext}$ sweet spot (where the first order derivative of frequency against external flux is zero) offers $T_\phi\gg T_1$ which becomes advantageous in the conversion of amplitude damping to an erasure. Specifically, we focus on the Integer Fluxonium Qubit (IFQ) externally biased at the zero flux sweet spot \cite{mencia2024integer,ardati2024bifluxon}, which can use a smaller bias current and a smaller flux loop to compensate for the environment flux bias. We denote the lowest three bare energy levels as $\ket{g}$, $\ket{e}$, and $\ket{f}$. The integer fluxonium has a highly coherent $\ket{g}-\ket{e}$ subspace. In this paper, we show that the $\ket{e}-\ket{f}$ subspace is also highly coherent due to the small splitting of the excited-state doublet.  

We study two alternative potential configurations of computational states of IFQ's bare energy levels: (1) the e-f qubit, which has the advantage of a suppressed transition frequency that protects from dielectric loss \cite{Mencia_thesis} and remotely resembles the superconducting dual-rail qubit since its computational states are symmetric and antisymmetric superpositions \cite{kubica2023erasure}. (2) The g-f qubit, which benefits from the forbidden direct transition between $\ket{g}$ and $\ket{f}$, enabling the erasure conversion of amplitude damping. The g-f qubit is related to the spin-locked g-f transmon in \cite{kubica2023erasure} and the g-f transmon ancilla used in a recent dissipative cat experiment \cite{AWS2024Cat_erasure}. We numerically demonstrate that both qubits are compatible with high-fidelity single-qubit operations.

To convert leakage to erasure without reducing the coherence of computational subspaces for the two IFQ configurations, we consider a modified form of dispersive readout. While dispersive readout is most commonly used to distinguish computational states and to destroy their coherence, here we aim to maintain coherence within the computational subspace while distinguishing it from an erasure state. We analytically derive the dephasing rate due to photon decay and phase smearing (when the qubit frequency is spread over a photon distribution). We then propose a bright-dark readout using a resonator that utilizes the fluxonium selection rules to map the lowest three levels of the IFQ onto two outcomes: (a) a bright resonator state with a large photon number indicating an erasure error or (b) a dark resonator state indicating the fluxonium state has remained in the computational subspace. The relatively lower photon number in the resonator during readout, when coupled to the computational states, reduces dephasing caused by phase smearing. This helps avoid introducing excessive error into the computational states during repeated erasure detection.

We show a geometric CZ gate between two g-f qubits is possible with direct charge coupling. Unlike geometric CZ gate in fluxonium-transmon-fluxonium (FTF) designs \cite{PhysRevX.11.021026,Ding2023FTF,qmds-z7gb} that use strong level repulsion from hybridization with a coupler to provide selectivity, we make up for weak frequency difference using the conditional frequency shift by using the selective darkening technique\cite{dogan2023CR,PhysRevApplied.18.034063,lin202424,de2010selective,de2012selective,PhysRevLett.127.130501,PhysRevLett.129.060501,jurcevic2021demonstration,PhysRevA.105.012602} to enhance selectivity. We further show high fidelity can be achieved at a gate time of $150$ ns by numerically optimizing the pulse, which enhanced the darkening of the unwanted, near-degenerate transition and gave the pulse envelope leakage removal features.

The paper is organized as follows. In Section \ref{sec:general_properties}, we introduce the properties of each proposed qubit and briefly discuss how to operate them. In Section \ref{sec:detection}, we discuss how to use dispersive readout for erasure detection while minimizing the amount of decoherence within the computational subspace. In Section \ref{sec:CZ} we present the direct charge coupling geometric CZ gate between two g-f qubits. 

We would like to bring the reader's attention to recent experimental work demonstrating erasure conversion and single qubit gates in the g-f qubit by Liu \textit{et al.}~\cite{liu2026convertingqubitrelaxationerasures} and An \textit{et al.}~\cite{an2025erasure,an2026erasure}. 

\section{The \MakeLowercase{e-f} and \MakeLowercase{g-f} qubit} \label{sec:general_properties}

\subsection{g-e-f three-level manifold}

The fluxonium is a single-mode superconducting circuit formed by a small Josephson junction with energy $E_J$ shunted by an inductance $L$ with energy $E_L=(\hbar/2e)^2/L$, and a capacitance $C$ with energy $E_C=e^2/2C$ (Fig.~\ref{fig:fig1} (a)). The circuit has the Hamiltonian
\begin{equation}
    \mathcal{H}_f = 4E_C\hat{n}^2  +\frac{1}{2}E_L\hat{\varphi}^2  -E_J\cos(\hat{\varphi}+\varphi_\text{ext}),
\end{equation}
where $\hat{n}$ is the Cooper pair number operator, $\hat{\varphi}$ is the phase difference operator, and $\varphi_{\rm ext}=2\pi \Phi_{\rm ext}/\Phi_0$ is the external flux bias through the fluxonium loop normalized to the flux quantum $\Phi_0 = h/2e$.

Following previous studies on IFQ \cite{mencia2024integer}, the parameter regime requires the localization of the three lowest eigenstates in distinct potential minima at $\varphi_\text{ext}=0$ and is achieved when
$E_L \ll E_J$, and $\sqrt{8E_J E_C}\gg 2\pi^2 E_L$.
Under these conditions, the ground state localizes in the central minimum ($m=0$), while the first two excited states are initially degenerate and confined to the adjacent minima ($m=\pm1$). Fluxon tunneling between these wells lifts the degeneracy, yielding symmetric and antisymmetric eigenstates, $\ket{e}$ and $\ket{f}$, respectively (see Fig.~\ref{fig:fig1}(b)).

Despite the relatively high transition frequency between $\ket{g}$ and $\ket{e}$, the corresponding charge matrix element is exponentially suppressed due to weak tunneling of the ground state into the two adjacent wells. In contrast, for the initially degenerate $\ket{e}$ and $\ket{f}$ states, the charge matrix element is also exponentially suppressed due to the transition frequency scaling (see Appendix \ref{sec:analytical}, eq.\ref{eq:frequency}, and IFQ matrix elements are also illustrated in Fig.~\ref{fig:fig1} (c)). Moreover, because of the symmetry of the potential, the transition between $\ket{g}$ and $\ket{f}$ is forbidden. This overall suppression of the charge matrix element for all allowed transitions effectively decouples the qubit from energy-loss mechanisms such as dielectric loss, suggesting that this three-level system should exhibit long bit-flip lifetimes. Furthermore, since the levels are first-order insensitive to $1/f$ flux noise, an enhanced pure-dephasing time is also expected, indicating the three-level structure of IFQ is a prime candidate for implementing erasure-error correction. 

While the IFQ was originally investigated as a conventional qubit utilizing the two lowest energy states, we propose that choosing a computational subspace formed by the higher excited states—specifically either the e-f or g-f manifold—can also yield a highly coherent qubit. In these alternative configurations, the energy level structure requires the detection of leakage, but their high coherence could offer advantages for quantum error correction. Crucially, each choice of computational subspace requires a distinct regime of circuit parameters ($E_J$, $E_C$, $E_L$) to maximize computational subspace coherence times and enable high-fidelity gates. Consequently, a single physical device is not intended to switch between the e-f and g-f configurations. Instead, in our numerical calculations, each set of fluxonium parameters is chosen exclusively for the respective protocol of that configuration. The e-f and g-f qubit operations are thus simulated and analyzed using the two distinct, optimized parameter sets detailed in Table~\ref{tab:qubit_parameters}. Note that, in this paper, we selected circuit parameters for the e-f and g-f qubits, so we can be specific in numerical simulations. There are continuous regions in the circuit parameter space where the two configurations can yield high coherence in their computational subspaces.

\begin{table}
\caption{\label{tab:qubit_parameters}Circuit parameters for the two example fluxoniums and their erasure detection configurations in GHz units. Note that we intend to use different circuit parameters for e-f and g-f computational subspaces because these two configurations have different parameter regime where they achieve longest effective coherence times.}
\begin{ruledtabular}
\begin{tabular}{cccccccccccccccccccc}
qubit& $E_J/h$&$E_C/h$& $E_L/h$& $\omega_q^{ge}/2\pi$& $\omega_q^{ef}/2\pi$  \\
\hline
e-f&  3   &  0.75    &   0.146 & $2.64$ & $0.03$  \\
g-f&   4  &  2   &      0.133 & $2.42$ &$0.135$ \\
\end{tabular}
\end{ruledtabular}
\end{table}

\subsection{Operating the e-f qubit}\label{subsec:ef_Xgate}

Utilizing the e-f manifold as the computational subspace benefits from a suppressed charge matrix element due to the low operational frequency, optimally ranging from tens to several hundreds of MHz. The decay from $\ket{e}$ to $\ket{g}$ brings the population out of the computational subspace, and we intend to convert that into the erasure error. Since the e-f transition has a low frequency, thermal excitation can cause $\ket{e}$ to excite into $\ket{f}$. The higher g-e frequency prevents the leaked population from spontaneously returning to the computational subspace, causing errors that are harder to correct than erasure errors. The idling error structure of the e-f qubit is illustrated in Fig.~\ref{fig:fig1} (d). 

In this paper, we use an example e-f qubit with $\omega_{ef}/2\pi=30 \text{ MHz}$ with the parameters specified in Table.~\ref{tab:qubit_parameters}. We show in Appendix \ref{ef_coherence} that this frequency is roughly optimal for maximizing the $T_2$ of the e-f subspace. A lower frequency would suffer from shorter $T_\phi$ due to the sweet spot having high curvature leading to significant decoherence from second-order flux noise, or having $T_1$ limited by $1/f$ flux noise in the heavy fluxonium regime of several $\text{MHz}$.

The small charge matrix element that arises in the relatively low-frequency computational subspace poses challenges for population transfer, which is not only required for gate operations but also for restoring the erasure population to the computational subspace for the g-f configuration. The low frequency of the e-f subspace makes flux driving a natural way to operate this qubit. Similar to \cite{campbell2020CQB,zhang2021universal}, flux pulses can drive the fluxonium away from the sweet spot to implement quantum gates on the qubit. These gates are applicable to the e-f qubit as long as the computational subspace remains intact after the gate. Regarding the erasure error channel, temporarily leaving the sweet spot allows the state $\ket{f}$ to decay to $\ket{g}$, resulting in slightly unbiased erasure. Since unbiased erasure is still easily correctable, this is not a major concern. Simultaneous flux and charge driving can also be utilized for fast high-fidelity qubit control \cite{MIT_counter_rotating}. 

\begin{figure}
    \centering
    \includegraphics[width=1.0\linewidth]{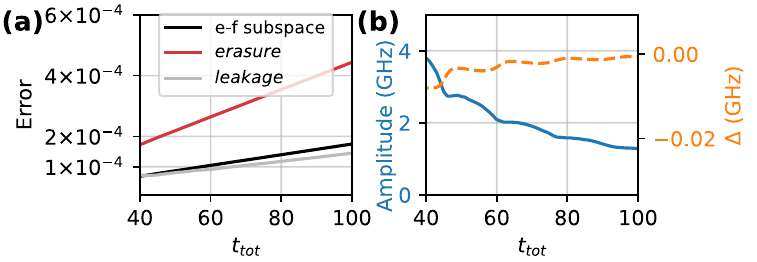}
    \caption{\textbf{(a)} Total gate time $t_\text{tot}$ of single qubit X gate on the example e-f qubit 
    versus error rates. Erasure error to state $\ket{g}$ dominates the total error. As the gate becomes shorter, leakage out of the g-e-f manifold becomes larger than the error within the computational subspace. \textbf{(b)} Optimized amplitude (which is not monotonic because $\sin^2$ ramp-up/down ratios (not shown) vary), and detuning $\Delta$ for the e-f qubit X gate.
    }
    \label{fig:ef_gate}
\end{figure}

A fast microwave pulse on a low-frequency qubit is still possible. In Fig.~\ref{fig:ef_gate}, we show the error metrics of a microwave activated X gate with drive term $H_\text{drive}=A(t)\sin(\omega_{ef}+\Delta)\hat{n}$, where $A(t)$ describe a square pulse with symmetric $\sin^2$ ramp-up/down, at different total gate time $t_\text{tot}$, obtained with optimized amplitude, detuning and ramp-up/down times. The noise model and the method for calculating the error rates are outlined in Section \ref{sec:noise_model} and \ref{fidelity_measure}. We observe that the renormalized computational subspace after truncation can achieve high fidelity provided the gate is longer than one Larmor period. The primary source of error is the erasure of the state $\ket{g}$. However, in our simulation with 20 levels, off-resonant excitation of levels outside the g-e-f manifold occurs on the same order of magnitude as computational subspace error, due to the large drive amplitude, and the leaked population can be stuck at higher levels at the end of the pulse due to decay. If additional leakage detection on those other levels is available, then leakage outside the g-e-f subspace may be less concerning in quantum error correction settings. It is worth mentioning that in this regime where gate time is not much longer than Larmor period, commensurate gates \cite{MIT_counter_rotating} and leakage cancellation \cite{4kz9-w97h} will be useful in actual implementation.

\subsection{g-f qubit single qubit Raman X gate}\label{subsec:g-f raman}
\begin{figure*}
    \centering
    \includegraphics[width=1.0\linewidth]{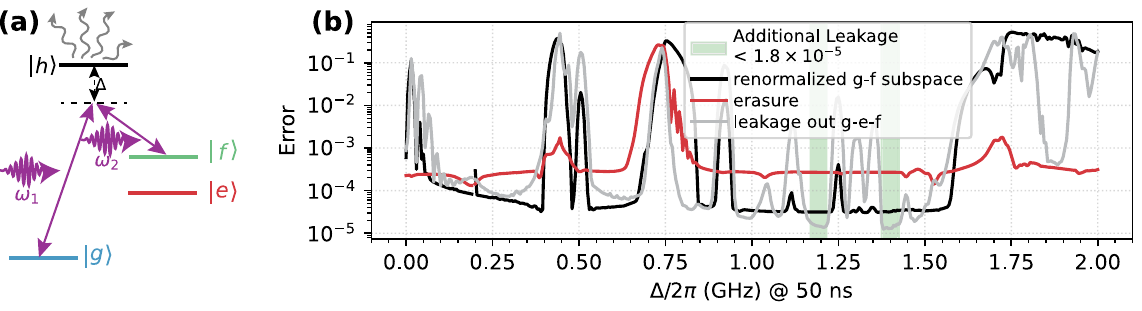}
    \caption{\textbf{(a)} Illustration of the g-f qubit Raman gate and its error structure. By applying two drives at frequencies $\omega_1=\omega_{gh}-\Delta, \omega_2=\omega_{fh}-\Delta$ at a relatively large detuning $\Delta$, the decay from the intermediate state $\ket{h}$ is minimized. The large drive amplitude induces some off-resonant excitation into higher levels, which then decay and become trapped in levels outside the computational subspace. The dominant error is decay from $\ket{f}$ to $\ket{e}$. This error is qualitatively equivalent to idling and can be detected via dispersive readout. \textbf{(b)} Error rates of the Raman gate optimized at different detunings, at a fixed total gate time $T_{gate}=50$ ns. Bumps in the error rate indicated unwanted single- or multi-photon resonances with certain transitions. For example, at $0.45$ GHz detuning, an unwanted $\ket{g}\rightarrow\ket{h}\rightarrow\ket{4}$ is resonant, at $0.7$ GHz, the second drive is in resonance with $\ket{g}\rightarrow\ket{e}$. At good detuning choices, such as $1.4$ GHz, the majority of the error is decay that is convertible to erasure (red line); the error of the re-normalized computational subspace (black line) is an order of magnitude lower; and the leakage outside the g-e-f subspace (grey line) is even lower.  }
    \label{fig:gates}
\end{figure*}
In the g-f qubit, direct bit-flips between the computational states $\ket{g}$ and $\ket{f}$ are symmetry-forbidden for the majority of loss mechanisms. Therefore, the decay from $\ket{f}$ to $\ket{e}$ can be converted into an erasure error for the g-f qubit. Cascaded bit-flips from $\ket{f}$ to $\ket{e}$ to $\ket{g}$ can be suppressed by prompt erasure conversion (see subsection \ref{sec:how_often}). The high g-e transition frequency ensures that $\ket{g}$ is immune to thermal excitations. The only remaining error in the computational subspace is dephasing. This idling error structure is illustrated in Fig.~\ref{fig:fig1} (e). 

The example g-f qubit we use in this paper has $\omega_{ef}/2\pi=0.134 \text{ GHz}$ with parameters specified in Table.~\ref{tab:qubit_parameters}. This relatively high $\omega_{ef}$ ensures the $T_\phi$ of the computational subspace to be sufficiently long for a high erasure ratio. This frequency is still low enough such that the erasure error which is related to $\Gamma_1^{ef}$ is still suppressed in absolute magnitude not to inundate the erasure conversion and error correction.

Similar to many metastable qubits, the vanishing matrix element in the g-f subspace makes direct gate operations challenging. Instead, a two-photon process based on an external ``$\Lambda$" structure is required for population transfer \cite{Henriet2020quantumcomputing,raman_qubit,zero_pi,t1protection}. For the g-f qubit, this ``$\Lambda$" structure can be realized as $\ket{g} \leftrightarrow \ket{h} \leftrightarrow \ket{f}$ where $\ket{h}$ is the third excited state. However, introducing intermediate states alters the error structure during gate operations, as decay from these states often reduces gate fidelity. By employing sufficiently large detuning from the intermediate state and avoiding unwanted resonances, the ideal error structure can be approximately maintained. 

In Fig.~\ref{fig:gates} (a), we summarize the concept of the Raman X gate for the example g-f qubit with parameters listed in Table \ref{tab:qubit_parameters}. In Fig.~\ref{fig:gates} (b), we use this simulation result to show that by making a good choice of the detuning to avoid single or multi-photon resonance with other unwanted transitions, error within the computational subspace can be reduced to about $5\times10^{-5}$ while the erasure error is around $4\times10^{-4}$. We use the noise model in \ref{sec:noise_model} and the method in Section \ref{fidelity_measure} to calculate the error rates. More details of the Raman gate are discussed in Appendix \ref{raman_details}.

\section{Converting decay to erasure via dispersive readout}\label{sec:detection}

To demonstrate the feasibility of erasure conversion for the e-f and the g-f qubits, we propose an erasure detection protocol based on dispersive readout. Here, a resonator is capacitively coupled to a fluxonium qubit. Upon being driven, the transmitted signal is used to distinguish erasure leakage states from computational states (Fig.~\ref{fig:fig1} (a)). We pay special attention to how we couple the resonator to the qubit, and how we drive the resonator to reduce dephasing in the qubit computational subspace.

\subsection{Fluxonium dispersive readout}

The Hamiltonian of the system is expressed as \cite{krantz2019guide}:

\begin{align}\label{hamiltonian}
&\mathcal{H} = \mathcal{H}_{f} + \omega_r \hat{a}^\dagger \hat{a}  + ig \hat{n}(\hat{a}^\dagger - \hat{a})
\end{align} 
Here, $\omega_{r}$ is the bare resonator frequency, $\hat{a}$ is the resonator annihilation operator $\hat{n}$ is the fluxonium charge operator and $g$ is the qubit-resonator effective coupling strength. Since the protocol requires precise tuning of $\omega_r$, we assume that the resonator is tunable.

In the dispersive regime(where coupling strength $g$ is small compared to transition detuning), a qubit in state $\ket{j}$   introduces the state-dependent shift to the resonator frequency $\chi_{j}$  \cite{zhu2013fluxoniumshift,blais2021circuitQED}:

\begin{equation}\label{eq:chi}
\chi_{j}=2g^2\sum_{i} \frac{(\omega_j-\omega_i)|\bra{i}\hat{n}\ket{j}|^{2} }{(\omega_j-\omega_i)^2-\omega_{r}^2}
\end{equation}
where $\omega_i$ and $\omega_j$ are the bare qubit frequencies corresponding to states $i,j$. Unlike a transmon qubit, multiple fluxonium transitions couple to the resonator due to the selection rules. Thus, the dispersive regime requirement $\Delta \gg g$ is mostly relevant to the fluxonium transition that strongly hybridizes with the resonator.

When the resonator is driven at frequency $\omega_d$ with amplitude $A$, the photon field $\alpha=\langle\hat{\alpha}\rangle$ inside the resonator evolves according to a semi-classical equation of motion dependent on the dispersive shift $\chi_j$:

\begin{equation}\label{eq:eqn_of_motion}
\dot{\alpha} = -i (\omega_{r}+\chi_{j}-\omega_d) \alpha - i A - \frac{\kappa}{2} \alpha,
\end{equation}
where $\kappa$ is the resonator's decay rate. This photon field is then detected and converted into classical signals for processing. 

To distinguish the signal from two different states $\ket{a}$ and $\ket{b}$, we can define the signal-to-noise ratio (SNR) from Hamiltonian simulation in terms of the Husimi-Q function $Q_\text{raw}$ of the reduced resonator density matrix $\rho_{res}$ by partial-tracing out the qubit state from the averaged density matrices, after changing to a product basis using the method in \cite{ionization}. $Q_\text{raw}$ approximates the amplified signal with vacuum noise \cite{blais2021circuitQED,caves2012qfunction}, which maps a complex value $\beta$ to a real value by $Q(\beta)=\bra{\beta}\rho_{res}\ket{\beta}$. Note that we re-scale the quadrature to let the center of the Q-function be consistent with the scale of the complex value $\alpha$, as shown in Fig.~\ref{fig:ef_readout} (b) and Fig.~\ref{fig:gf_readout} (b). The signal formed by photon decay to the transmission line is then scaled by $\sqrt{\kappa}$: $Q(t) = \sqrt{\kappa}Q_\text{raw}(t)$. The signal after integration over a time period $\tau$ is therefore:
\[
M = \sqrt{\kappa}\int_{0}^{\tau} w(t)Q(t) dt
\]
where $w(t)$ is the weighing function that normalizes to $\tau$: $\int_{0}^{\tau} w^2(t)dt = \tau$. To maximize contrast, we use a rotating projection axis of angle $\theta$ to minimizes the overlap between the marginal distributions of $Q$ on that axis (Radon transform of $Q$ with angle $\theta$). These marginal distributions are well-approximated by 1-D Gaussians with parameters $\mu_a(t), \mu_b(t),  \sigma^2_a(t), \sigma^2_b(t)$. We take the weighing function to be proportional to the contrast of the Q functions: $w(t)\propto |\mu_a(t) - \mu_b(t)|$ as in \cite{walter2017single_shot}. Thus we have \cite{walter2017single_shot}
\begin{align}\label{eq:qfunc_accuracy}
&\text{SNR}(\tau) = \\
&\sqrt{\kappa}\frac{\int_{0}^{\tau} w(t)|\mu_a(t)-\mu_b(t)|dt}{|\int_{0}^{\tau}  w(t)\sigma^2_a(t)dt+\int_{0}^{\tau}  w(t)\sigma^2_b(t)dt|^{1/2}}.  \nonumber
\end{align}
Finally, the measurement assignment error can be computed from the SNR as
\[
\text{error} = \text{erfc}\Big(\frac{\text{SNR}}{2}\Big).
\]

\subsection{Readout induced dephasing}

In addition to estimating the distinguishability of signals from the leakage state, we are also concerned with not distinguishing between the two computational states. We are thus interested in modelling the amount of dephasing error introduced into the computational subspace during readout. We model the dephasing error as a resonator decay-induced part and a decay-independent part. In Appendix \ref{sec:dephasing_incoherent}, we derive the qubit exponential decoherence rate due to resonator decay to be:  
\begin{equation}\label{measurement_rate}
    \begin{aligned}
\Gamma_\varphi^{readout}= \frac{2 A^2 \kappa(\chi_a -\chi_b)^2 }{(4(\Delta_{dr}+\chi_a)^2+\kappa^2)(4(\Delta_{dr}+\chi_b)^2+\kappa^2)},
    \end{aligned}
\end{equation}
where $\Delta_{dr} = \omega_r - \omega_d$ is the detuning of the drive from the bare resonator frequency. 

The second, decay-independent part is related to the smearing effect of phase spread. Assuming a linear photon number dependence of frequency $\omega_q(n) = \omega_0 + n \Lambda$, where $\Lambda$ is the AC-Stark shift, and assuming a Poisson distribution of photon number, the averaged phase accumulation of the qubit, as described by the off-diagonal matrix element, is 
\[
\langle\rho_\text{off-diag}(t)\rangle = \rho_\text{off-diag}(t=0)e^{i\omega_0 t}\langle e^{i\Lambda n t}\rangle,
\] 
and the amplitude of the average explains the dephasing $|\langle e^{i\Lambda n t}\rangle|
=\exp\left(-\Gamma(t)\right)$, which can be written as  
\begin{equation}\label{eq:smearing_equation_combined}
\Gamma(t)=
\Lambda^2
\int_{0}^{t}
\!\!\int_{0}^{t}
\sqrt{\overline{n}(t_1)\,\overline{n}(t_2)} \,
\exp\!\bigl[-\kappa|t_1 - t_2|\bigr] \,dt_1\,dt_2,
\end{equation}
where $\bar n(t)$ is the average photon number of the coherent state as a function of time, which we can approximate as $\overline{n}(t) = n_{max}\sin^2(\pi t / t_{max})$ using the semi-classical equation of motion eq.\eqref{eq:eqn_of_motion}, or from the black curves in Fig.~\ref{fig:ef_readout} (d) and Fig.~\ref{fig:gf_readout} (d)(see Appendix \ref{sec:dephasing_smearing} for derivation of Eq \eqref{eq:smearing_equation_combined}).

\begin{figure}
    \centering
    \includegraphics[width=1.0\linewidth]{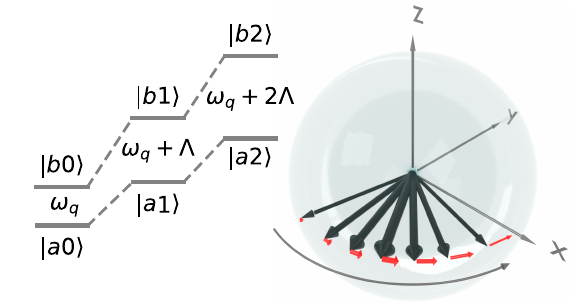}
    \caption{ Illustration of dephasing caused by photon number dependent qubit frequency shift, or AC-Stark shift. In a simplified model good enough to model our leakage detection, the amount of dephasing is proportional to the square of photon number frequency dependency $\Lambda$ (Eq. \eqref{eq:smearing_equation_combined}).}
    \label{fig:nonlinearity}
\end{figure}

\subsection{Using dispersive readout for erasure detection}\label{subsec:dispersive_detection}

Now, without loss of generality, let $\ket{a}$ and $\ket{b}$ be the two computational states from which we want to distinguish an erasure state $\ket{L}$. We do not want the erasure detection to distinguish $\ket{a}$ from $\ket{b}$ because that means destroying the coherence within the $\ket{a}$-$\ket{b}$ subspace. Therefore, we want to find the resonator frequency $\omega_r$ that satisfies 
\begin{equation} \label{eq:chi_condition}
\chi_L \neq \chi_a, \quad  |\chi_a - \chi_b| \approx 0.
\end{equation}

In many readout protocols, the two computational states are exposed to a large number of photons \cite{Jay2008readout,longitudinal}. The large average photon number and the photon-number dependence of the qubit frequency lead to dephasing in the computational subspace. We are thus motivated to minimize the photon number in the resonator when coupled to computational states. 

\begin{figure}
    \centering
    \includegraphics[width=1.0\linewidth]{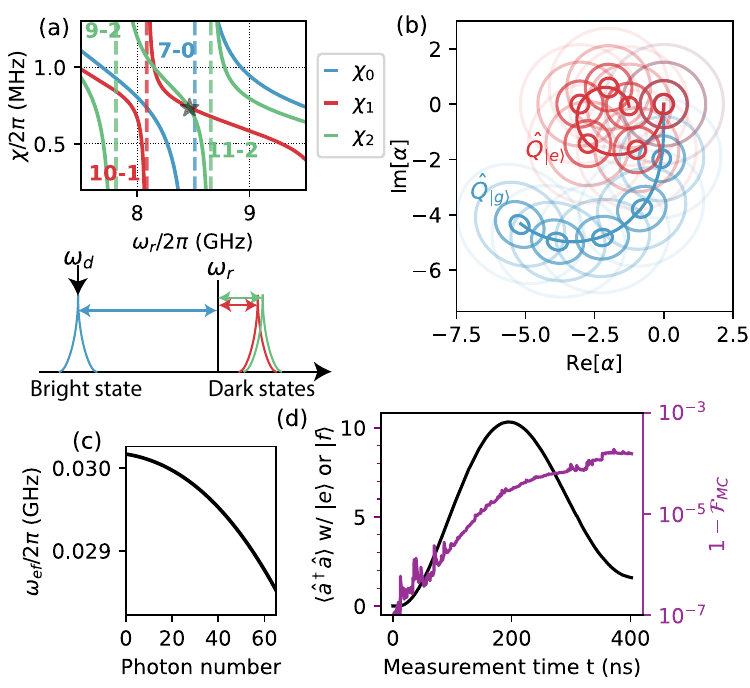}
    \caption{\textbf{(a)} For the example e-f qubit, we utilize the $\ket{0} - \ket{7}$ plasmon transition to induce large $\chi_0$. In the meantime, $\chi_1\approx\chi_2$ are small because their dispersive shifts are affected by weak transition elements. We drive at the resonator frequency dressed by $\ket{g}$, which is $2\pi \times 2.5 $ MHz detuned from the frequencies dressed by $\ket{e}$ or $\ket{f}$. (Plot generated from perturbation theory \cite{zhu2013fluxoniumshift}. See table \ref{tab:chi} for dispersive shifts, coupling strengths, and resonator frequencies from numerical diagonalization that we use in simulation.)
    \textbf{(b)} Evolution of the coherence state $\alpha$ from Monte Carlo simulation coupled to the example e-f qubit. The Husimi-Q function is shown as contours that encloses $0.1,0.5,0.9,0.99$ of the cumulative probability density. The coherent state trajectory for state $\ket{f}$ almost completely overlaps with the other computational state and is thus not drawn. The blue color stands for the fluxonium state $\ket{g}$, and the red color stands for the state $\ket{e}$. The resonator is driven at $A = 0.063$ GHz, and has photon lifetime $1/\kappa=2\times10^{-7}$ s.
    \textbf{(c)} AC-Stark shift of the example e-f qubit. It is not entirely linear, which means Eq. \eqref{eq:smearing_equation_combined} does not accurately predict the amount of dephasing error. 
    \textbf{(d)} The photon number in the resonator when coupled to computational states (black) and the qubit computational subspace error rate (purple) during the dispersive detection on example e-f qubit as a function of integration (measurement) time. The qubit error is computed by evolving the tomographic qubit states coupled with the resonator: \( y_0 = \{\pm\ket{X}, \pm\ket{Y}, \pm\ket{Z}\} \otimes \ket{0} \).}
    \label{fig:ef_readout}
\end{figure}

\begin{figure}
    \centering
    \includegraphics[width=1.0\linewidth]{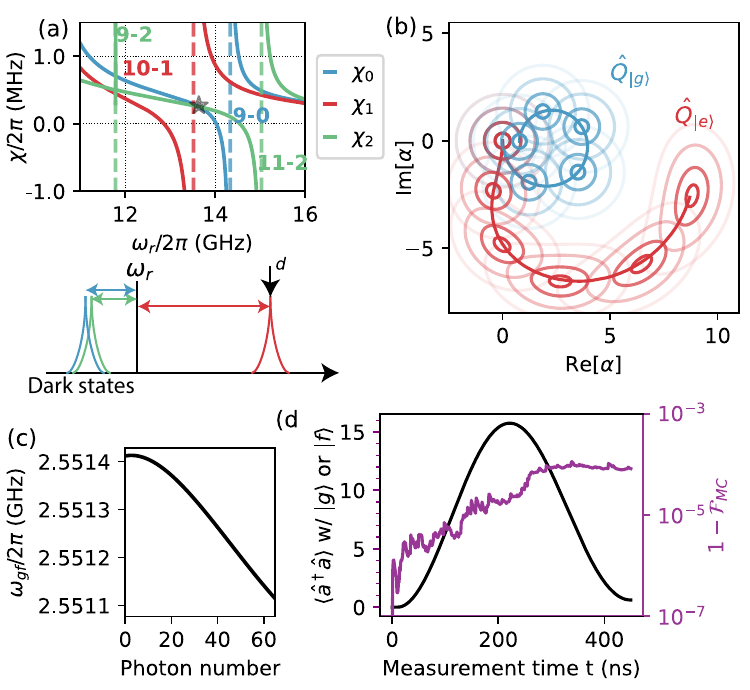}
    \caption{\textbf{(a)} To detect leakage to state $\ket{e}$ in the example g-f qubit, we use the $\ket{e}-\ket{10}$ plasmon transition to provide a large $\chi_{e}$. The on-resonance $\ket{f}-\ket{10}$ transition is parity forbidden. Other closely detuned transitions like $\ket{f}-\ket{9}$, $\ket{f}-\ket{11}$, $\ket{0}-\ket{9}$, $\ket{0}$ and $\ket{2}$ collectively contribute to $\chi_g \approx\chi_f$ by numerical $\chi$ matching. We drive at the resonator frequency dressed by $\ket{e}$, which is $2\pi \times 3.4 $ MHz detuned from the frequency dressed by the computational states. (Plot generated from perturbation theory \cite{zhu2013fluxoniumshift}. See table \ref{tab:chi} for dispersive shifts, coupling strengths, resonator frequencies from numerical diagonalization that we use in simulation.) \textbf{(b)} Evolution of the coherence state $\alpha$ driven at $A = 0.063$GHz, and has photon lifetime $1/\kappa=5\times10^{-7}$ s. \textbf{(c)} AC-Stark shift of the example g-f qubit. \textbf{(d)} Same as Fig.~\ref{fig:ef_readout} (d)}
    \label{fig:gf_readout}
\end{figure}

We take inspiration from photon-counting readout protocols \cite{nesterov2021countingstatistics}, where the qubit states are assigned as either bright or dark. The resonator is driven at the frequency dressed by the qubit state designated as the bright state: $\omega_d = \omega_{r} + \chi_{\text{bright}}$. Compared to the traditional choice of drive frequency, the bright-dark readout reduces the number of photons in the resonator when coupled to the dark state, as $\omega_d$ is twice as detuned (this reduction in photon number can be see from eq.\eqref{eq:eqn_of_motion}. We illustrate the choice of drive frequency regarding dispersive shift in Fig.~\ref{fig:ef_readout} (a) and Fig.~\ref{fig:gf_readout}(a). As in \cite{nesterov2021countingstatistics}, we also choose the drive time to be approximately an integer multiple of $\pi/(|\chi_{\text{bright}}|+|\chi_{\text{dark}}|)$, so the photon number returns to nearly zero when coupled to the dark state, according to the semi-classical eq.\eqref{eq:eqn_of_motion}.

\subsection{Simulation results}

\begin{table}
\caption{\label{tab:chi}The erasure detection configurations and dispersive shifts in the simulation for the two example fluxoniums (Unit: GHz).}
\begin{ruledtabular}
\begin{tabular}{cccccccccccccccccccc}
 qubit  &g&  $\omega_r/2\pi$  & $\chi_g/2\pi$  &  $\chi_e/2\pi$ & $\chi_f/2\pi$\\
\hline
e-f& 0.25   &8.461 & $-1.57 \times 10^{-3}$ & $0.923 \times 10^{-3}$ & $0.919 \times 10^{-3}$  \\
g-f& 0.3 &  13.634  & $0.378 \times 10^{-3}$ & $3.82 \times 10^{-3}$ & $0.379 \times 10^{-3}$  \\
\end{tabular}
\end{ruledtabular}
\end{table}

Condition \eqref{eq:chi_condition} can be satisfied in the IFQ by carefully selecting the resonator frequency. (This requirement may require a frequency-tunable resonator in actual hardware implementation to achieve precise frequency targeting.) For the example qubits, we find such resonator frequencies and list the parameters in TABLE \ref{tab:chi}. Although the resonator frequencies in our numerical examples are relatively high, they be reduced by proportionally scaling down the fluxonium parameters $E_J$, $E_C$, and $E_L$. For the e-f qubit, we utilize the relatively large $\ket{g} - \ket{7}$ transition element to induce large $\chi_g$ while keeping $\chi_e\approx\chi_f$ small by keeping only small transition elements from $\ket{e}$ and $\ket{f}$ near-detuned (the 2-11 and 1-10 transition induces narrower poles in Fig.~\ref{fig:ef_readout} (a)). For the example g-f qubit, we use the plasmon $\ket{1} - \ket{10}$ transition for large $\chi_e$ while using the combined effect of selection rule and parameter matching for $\chi_g\approx\chi_f$ (Fig.~\ref{fig:gf_readout} (a)). 

The resonator frequencies, coupling strengths, and drive frequencies are listed in Table \ref{tab:qubit_parameters} and illustrated in Fig.~\ref{fig:ef_readout} (a), \ref{fig:gf_readout} (a). We drive the resonator at amplitude $A=0.0628\text{ GHz}$ in both cases. The evolution of the coherent state under drive is plotted in Fig.~\ref{fig:ef_readout} (b) and Fig.~\ref{fig:gf_readout} (b). For both qubits, we obtain an SNR of about 7 (Fig.~\ref{fig:ef_readout} (d) and Fig.~\ref{fig:gf_readout} (d)), which results in readout assignment errors of $7\times10^{-7}$. Because we have assumed a readout efficiency of unity, a realistic readout efficiency number can increase the error in actual experiments. This simulated accuracy is good enough as a preliminary study and still shows the viability of erasure leakage detection in IFQ when condition \eqref{eq:chi_condition} is satisfied. In principle, erasure check assignment errors can be easier to handle than undetectable gate errors within the computational subspace in quantum error correction decoders\cite{spatial_temporal_imperfection_of_erasure_checks,kubica2023erasure}

We observed error in the computational subspace at around $10^{-3}$ ($10^{-5}$) for the example e-f (g-f) qubit from the simulation (purple curves in Fig.~\ref{fig:ef_readout} (d) and Fig.~\ref{fig:gf_readout} (d)). We also simulated the evolution without resonator decay and observed even higher error rates. This suggests that the majority of errors are due to resonator-decay-independent dephasing (phase smearing arising from the photon-number dependence of the qubit frequency). This can be understood intuitively: Without resonator decay, the wider spread of photon number distribution causes more coherence to be lost when averaged together (the spread of the distribution is illustrated in Fig.~\ref{fig:nonlinearity}). The simplified model in \eqref{eq:smearing_equation_combined} matches the amount of dephasing observed in the simulation. Some small deviations from the master equation most likely come from the non-linearity of the AC-Stark shift $\Lambda$ (shown in Fig.~\ref{fig:ef_readout} (c) and  Fig.~\ref{fig:gf_readout} (c)). Since the phase smearing effect occurs coherently, a dynamical decoupling sequence or continuous driving might be able to mitigate it, potentially easing the stringent parameter-matching requirements when using a resonator for leakage detection.

\subsection{Prediction on requirements of $\chi$ matching}\label{subsec:predictions}

\begin{figure}[h]
    \centering
    \includegraphics[width=1.0\linewidth]{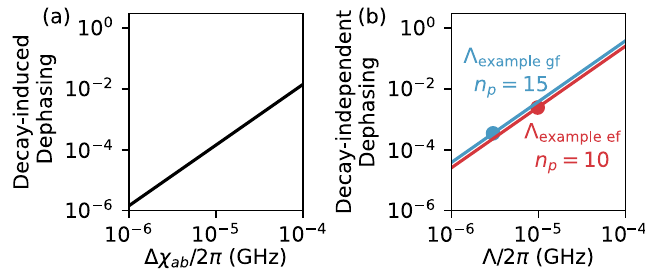}
    \caption{\textbf{(a)} The amount of dephasing caused by dispersive leakage detection in the qubit computational subspace as a function of difference in dispersive shift from computational states $\Delta\chi_{ab}$, or \textbf{(b)} as a function of photon number-dependent qubit frequency shift $\Lambda$. The calculations in (a) and (b) follow the assumptions mentioned in \ref{subsec:predictions}.}
    \label{fig:dephasing_estimate}
\end{figure}

With the analytical expressions \eqref{measurement_rate}, \eqref{eq:smearing_equation_combined}, we can now predict the amount of dephasing error caused by the two different mechanisms in isolation during the dispersive readout leakage detection. To calculate the amount of dephasing caused by resonator decay after $T_{meas}=400$ ns of readout (we picked 400 ns to be consistent with the readout for the example e-f qubit) in the steady state, we vary $\Delta\chi_{ab}$ (the unwanted mismatch of $\chi$ between the two computational states a and b) in the numerator in Figure \ref{fig:dephasing_estimate} (a), and compute this part of dephasing error assuming the same drive amplitude and resonator lifetime as in the e-f qubit readout example of $A=0.063 \text{ GHz}$, $1/\kappa=2\times10^{-7}$ s, and approximate the denominator with $\Delta_{dr}+\chi_a\approx\Delta_{dr}+\chi_b=2\pi \times 2 \text{ MHz}$ which means a reasonable $2$ MHz separation between the $\chi$ of computational states and the leakage state. For the decay-independent part, we assume the average photon number to follow $n(t)=n_p\sin^2(\pi t/T_{meas})$, and plot the dephasing rate as a function of photon number-dependent qubit frequency shift and $n_p$ in Figure \ref{fig:dephasing_estimate} (b). We observe that as long as the difference in the dispersive shift ($\Delta\chi_{ab}$), and the AC-stark shift $\Lambda$ is small enough, the amount of dephasing induced on the qubit can be negligible. (For reference, the dispersive shift of the two example setups is mentioned in the caption of Table \ref{tab:chi}, and the $\Lambda$ of the two example qubits are shown in Figure \ref{fig:ef_readout} (c) and Figure \ref{fig:gf_readout} (c). We calculate $\Lambda$ by using the slope of qubit frequency in the segment of the first 50 photons.)

\section{g-f qubit CZ gate}\label{sec:CZ}

Because the e-f qubit is very similar to heavy fluxonium, we focus only on the two qubit gate for the g-f qubit in this paper. Specifically, we discuss implementation strategies for a geometric CZ gate via higher levels.

Two qubit gates on g-f qubit should preserve the error structure of no direct dissipation between $\ket{g}$ and $\ket{f}$ states. Conditional population transfer between $\ket{g}$ and $\ket{f}$ states is a 2-photon non-obvious process, and we focus on conditional phase accumulation in this section. An obvious strategy is to dynamically shift the $\ket{ff}$ level energy in two coupled g-f qubits, similar to \cite{PhysRevApplied.20.044012,Nondegenerate,PhysRevA.87.052306,PhysRevB.81.134507}, and the techniques used in these references may apply to g-f qubits here. Another method to achieve a conditional phase is to use a geometric phase via a $2\pi$ round-trip transition \cite{PhysRevX.11.021026,Ding2023FTF,qmds-z7gb} from the $\ket{ff}$ state to and from another state, which we choose to implement using direct coupling in this paper.

\begin{figure}
    \centering
    \includegraphics[width=1\linewidth]{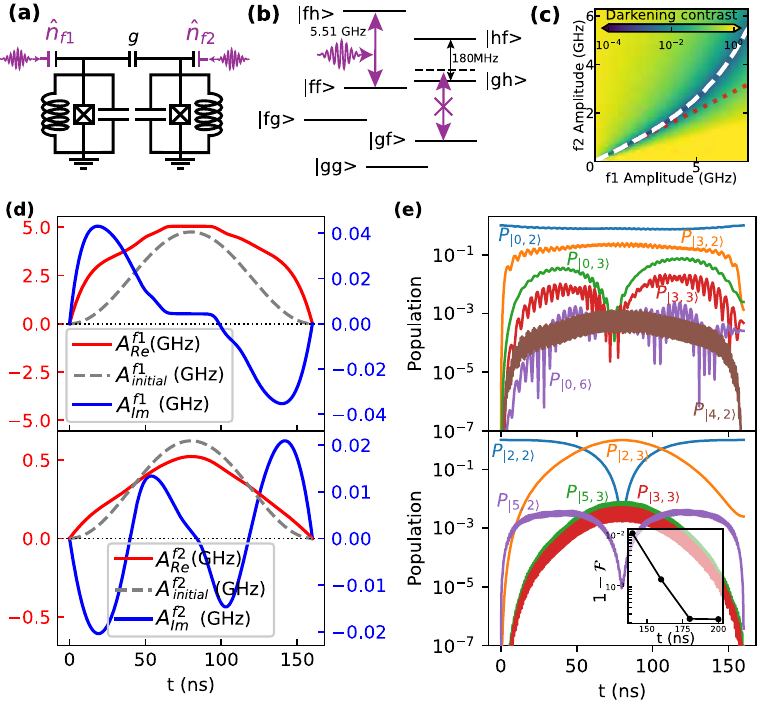}
    \caption{g-f qubit direct weak coupling CZ gate based on selective darkening. \textbf{(a)} Circuit diagram of two capacitively coupled fluxoniums, the waveform represent our scheme of driving the charge operator of both two qubits. \textbf{(b)} Simple geometric phase CZ gate via intermediate state $\ket{h}$. The $\ket{f}-\ket{h}$ transition of the second g-f qubit is $180$ MHz detuned from the $\ket{g}-\ket{h}$ transition of the first g-f qubit, which creates very small hybridization. However in this scheme the gate speed is still limited by leakage to strongly hybridized levels.\textbf{(c)} The heatmap shows that the instantaneous darkening ratio (dashed white curve) deviates from the static selective darkening ratio (represented by dashed red line) at large drive amplitudes. \textbf{(d)} The numerically optimized pulse that corrects for the nonlinear effects in large drives. The real ($A_{Re}$) and imaginary ($A_{Im}$) envelopes on the two qubits are shown in red and blue. The dashed gray curves represent the initial guess raised cosine envelope. \textbf{(e)} Population evolution when simulated without dissipative terms shows that the wanted $\ket{22}-\ket{23}$ process, and disables the unwanted degenerate process $\ket{02}-\ket{03}$. The inset of the bottom panel shows the optimized coherent fidelity as a function of gate time. Coherent fidelity at low digits of $10^{-4}$ can be achieved above 175 ns.}
    \label{fig:gf-2q}
\end{figure}

In the weak coupling regime, there is no large frequency difference between the wanted and unwanted transitions that provides the large detuning for selectivity in driving. To achieve conditional population transfer, we resort to the selective darkening technique used for CNOT gates \cite{dogan2023CR,PhysRevApplied.18.034063,lin202424,de2010selective,de2012selective,PhysRevLett.127.130501,PhysRevLett.129.060501,jurcevic2021demonstration,PhysRevA.105.012602}. In addition to driving the second qubit for the $\ket{ff}-\ket{fh}$ transition, we apply a second drive on the first qubit at the same frequency with amplitude satisfying
\begin{equation}\label{eq:darkening_ratio}
    A_{f1}\bra{gf}\hat{n}_{f1}\ket{gh}+A_{f2}\bra{gf}\hat{n}_{f2}\ket{gh} = 0.
\end{equation}
This condition fixes the ratio between the drives on qubits 1 and 2. The overall drive amplitude is chosen to achieve a $2\pi$ Rabi oscillation.

Strong direct coupling is dangerous because it risks unwanted population transfer. We proceed with weak hybridization by picking a second g-f qubit whose $\ket{f}-\ket{h}$ transition is $180$ MHz detuned from the $\ket{g}-\ket{h}$ transition of the first g-f qubit. This shifts the frequency of the $\ket{f}-\ket{h}$ transition of the second qubit when coupled to the $\ket{g}$ state of the first qubit very weakly (below $1$ MHz).    
Although the dispersive shift on $\ket{gh}$ is small (first index is for first g-f qubit, second index is for second g-f qubit), the hybridization of $\ket{gh}$ with the $\ket{hf}$ state causes the second fluxonium $\bra{gf}\hat{n}_2\ket{gh}$ matrix element to acquire a component of the bare $\ket{g}-\ket{h}$ transition of the first g-f qubit, and that bare transition is a plasmon transition with a large matrix element. Note that in this scheme, we have a minimal ZZ rate, because neither of the two g-f qubits' $\ket{g}-\ket{f}$ subspace is used to hybridize to create dispersive shifts or cross-matrix elements for this gate. 
The level diagram of this CZ gate is shown in Fig.~\ref{fig:gf-2q} (b). Qubit 1's parameters are listed in table \ref{tab:qubit_parameters}, and qubit 2 has $E_J/h=8$ GHz,$E_C/h=4.2$ GHz, $E_L=0.1$ GHz, coupled to qubit 1 with $g=0.4$ GHz.

Selective darkening was proposed in the strong-coupling regime for cross-resonance gates, where the ``cross" drive (the drive on the qubit that's not supposed to transition) should be the primary drive, and the direct drive on the qubit where the transition is desired should be small. Our setup lacks the nice matrix-element structures; as a result, both drives must be comparable to achieve the darkening effect, which makes the drive amplitudes very large for short gates and leads to significant leakage. This makes it impossible to implement fast gates with simple pulse shapes without mitigating leakage. 
A good ($0.9993$) fidelity with a raised cosine pulse is only obtained at a gate time of $1000$ ns. The fidelity in the CZ gate simulation is obtained by first computing the propagator of the $7\times7=49$ level Hamiltonian, truncating to the 4 dimensional subspace, sandwiching with single qubit phase gates and then using $$\mathcal{F}=\frac{1}{20}\left(\text{Tr}\{U_{\text{ac}}^\dagger U_{\text{ac}}\} + \left| \text{Tr}\{U_{\text{id}}^\dagger U^{}_{\text{ac}}\} \right|^2\right).$$ In this $1000$ ns gate, most of the error is due to the unwanted $\ket{gf}-\ket{gh}$ transition. We report that in a faster analytical raised cosine pulse, spurious excitation of the $\ket{g}-\ket{h}$ excitation of qubit 1 also becomes non-negligible. The failure of simple analytical pulse shapes suggests that the ratio computed from static matrix elements no longer holds at short gate times as nonlinear effects emerge. In Fig.~\ref{fig:gf-2q} (c), we show a numerical simulation where we instantaneously drive the system at two amplitudes and compute the darkening contrast between wanted and unwanted transitions. At high amplitude, the ratio that maximizes ``darkening" of the unwanted transition begins to deviate from the straight line representing equation \ref{eq:darkening_ratio}, and that the maximally achievable darkening contrast decreases as amplitudes get larger (The color is brighter in the dark valley in high amplitude regions even after accounting for the correction).

To dynamically trace the amplitude-dependent selective darkening ratio and reduce leakage effects, we express the pulse envelope (on the two qubits, in both phase and quadrature components) using a smooth shape ansatz and then numerically optimize it using the computational subspace coherent fidelity as the cost function. We obtain a coherent fidelity of $0.9989$ at $160$ ns, and $0.99976$ at $180$ ns. We show the optimized $160$ ns pulse envelope in Fig.~\ref{fig:gf-2q} (d) and the population evolution from $\ket{gf}, \ket{ff}$ initial states in Fig.~\ref{fig:gf-2q} (e). The evolution history shows a pronounced darkening of the unwanted $\ket{gf}-\ket{gh}$ transition, which is only $f_{fh-ff}-f_{gh-gf}=0.45$ MHz detuned from the driven transition, as well as suppression of other leakage levels. In fact, the optimized pulse shape exhibits features resembling those of analytical leakage-reduction pulse shapes, such as DRAG \cite{PRXQuantum.5.030353, PhysRevLett.103.110501}.

In Appendix \ref{sec:STIRAP} we also discuss a potentially viable strategy to use another state as the direct coupling geometric CZ intermediate state to circumvent the error resulting from the undetectable decay from state $\ket{h}$ to computational states that violate the erasure-dominated error structure.

\section{Summary}\label{discussion}

In summary, we examined two IFQ configurations, the e-f and g-f qubits, designed to leverage the fluxonium level structure to enhance computational coherence times compared to conventional fluxonium qubits. This approach effectively boosts coherence without requiring complex qubit architectures or composite qubits \cite{PRXQuantum.2.010339,kou2024erasurefluxonium,kubica2023erasure,kubica2023heralding}. 
With our conservative error model, single qubit $40$ ns X gate on the e-f qubit can achieve $10^{-4}$ error rate within the computational subspace and $3\times10^{-4}$ transition to the state $\ket{g}$, which corresponds to an erasure rate. While a single qubit Raman X gate on g-f qubit can achieve $5\times10^{-5}$ computational error and $4\times10^{-4}$ transition to erasure state $\ket{e}$ at $50$ ns. Both demonstrate erasure-dominant error structure.

We introduced an erasure conversion for the transitions to $\ket{g}$ from e-f qubit and $\ket{e}$ from g-f qubit. Through numerical simulations of dispersive readout, we show that the classification error is below $10^{-6}$. We showed that readout-induced dephasing in the computational subspace is well tolerated in the QEC context for small dispersive shift mismatch and small AC-Stark shift of the qubit frequency.

We discussed the trade-offs among dephasing rate, bit-flip rate, and erasure rate for the e-f and g-f qubits. While the e-f qubit can perform better than the half-integer fluxonium in certain scenarios, the g-f qubit promises order-of-magnitude effective coherence time improvement compared to conventional fluxonium, benefiting from (1) the erasure conversion protocol and (2) gates that preserve the erasure-dominated idling error structure.

Computational subspace measurement for the e-f and g-f qubits can be performed by mapping one of the computational states to leakage states and detecting the resulting leakage. The reset is then done via classical feedback. This approach could mitigate Purcell decay and thermal photon-induced dephasing that arise from the idling of a dedicated computational readout resonator. Three-outcome measurements that distinguish all three states may be achieved by tuning the resonator frequency and flux to maximize the separation of dispersive shifts \cite{stefanski2023fluxpulseassisted}, or by sequentially applying two two-outcome measurements. Another method for performing a reset in IFQ has already been addressed in \cite{mencia2024integer}. Techniques for conventional fluxonium reset, such as tuning away from the sweet spot \cite{off_sweet_spot_initialization_fluxnoium}, may also be applicable. 

Two-qubit gate operations on the e-f qubit should also be similar to the operations of low-frequency fluxonium at half flux quantum (HFQ), which include techniques such as using transition matrix elements outside the computational subspace \cite{dogan2023CR,xiong2022cphase} and inductive coupling \cite{Zhang2024inductive}. Alternatively, native parity measurements could be used in place of entangling gates \cite{nick2025}. For the g-f qubit, while we showed a geometric CZ gate with direct capacitive coupling can work with high-coherent fidelity, implementation using a coupler is also worth exploration. A potential candidate is the STIRAP-based CZ gate proposed in \cite{setiawan2023STIRAP}.

Using the computational subspace and the erasure error rates, one can assess the QEC performance of the e-f and g-f qubits. Future work may study how leaked qubits interact with other qubits, which will help evaluate the trade-offs between leakage detection via hardware additions and leakage detection via software that analyzes correlated error signals \cite{eraser}.

The g-e IFQ qubit configuration also presents an erasure-dominant error structure. By suppressing tunneling, $T_1^{ge}$ can, in principle, be significantly increased. Unlike the e-f qubit, $T_1^{ge}$ is immune to $1/f$ noise due to its higher frequency. The loss of $T_\phi^{ge}$ from a sharper avoided crossing can be mitigated through dynamical decoupling. The dominant error is then thermal excitation from $\ket{e}$ to $\ket{f}$, which can be detected as biased erasure errors. 

Given its significantly longer effective coherence time, compared to other single-mode superconducting qubits in the QEC context, and the absence of clear obstacles in implementing a gate set that preserves an erasure-dominant idling error structure, the g-f qubit emerges as qualitatively superior to previously studied single-mode qubits.

\section{acknowledgements}
This research was supported by the ARO GASP (contract No. W911-NF23-10093) program. We acknowledge the use of the QuTiP software package \cite{qutip1}. Numerical simulations were performed using the computational resources and assistance of the UW-Madison Center for High Throughput Computing (CHTC) in the Department of Computer Sciences. The CHTC is supported by UW-Madison, the Advanced Computing Initiative, the Wisconsin Alumni Research Foundation, the Wisconsin Institutes for Discovery, and the National Science Foundation, and is an active member of the Open Science Grid, which is supported by the National Science Foundation and the U.S. Department of Energy’s Office of Science. Part of this research was conducted while the author was visiting the Institute for Mathematical and Statistical Innovation (IMSI), supported by the National Science Foundation (Grant No. DMS-1929348).

\appendix
\renewcommand{\thesection}{\Alph{section}}
\setcounter{figure}{0}
\renewcommand{\thefigure}{\thesection\arabic{figure}}

\section{Analytically understanding the qubit}\label{sec:analytical}

In this section, we analyze the second-order derivative of the energy levels of the integer-fluxonium qubit and relate it to the trade-offs between flux-noise sensitivity and frequency suppression.

The effective triple-well Hamiltonian for integer fluxonium can be approximated as \cite{mencia2024integer}:
\begin{align}\label{hamiltonian:int}
    H_{\text{eff}}&=\begin{pmatrix}
\frac{E_L}{2}(-2\pi-\varphi_\text{ext})^2 &-\epsilon_1/2&\epsilon_2/2\\
-\epsilon_1/2&\frac{E_L}{2}\varphi_\text{ext}^2&-\epsilon_1/2\\
\epsilon_2/2&-\epsilon_1/2& \frac{E_L}{2}(2\pi-\varphi_\text{ext})^2 
\end{pmatrix}.
\end{align}
where $\epsilon_1$ and $\epsilon_2$ are single and double well tunneling amplitudes.

Setting $\varphi_\text{ext}=0$, the transition frequencies can be computed from eigenenergies: $\omega_{ge} = E_e-E_g \approx 2\pi^2E_L - \epsilon_2/2 + \epsilon_1^2/4\pi^2E_L$, and $\omega_{ef}=E_f-E_e= \epsilon_2+\epsilon_1^2/4\pi^2E_L$
For small deviations around the sweet spot $\varphi_\text{ext} = 0+\delta\varphi$, the perturbation first order in $\delta\varphi$ is:
\begin{align}
    V=\begin{pmatrix}
2\pi E_L \delta\varphi &0 & 0\\
 0& 0 & 0\\
0 &0 &  -2\pi E_L \delta\varphi
\end{pmatrix}.
\end{align}

\textbf{Frequency and curvature:} Using non-degenerate perturbation theory, the second order correction to eigenenergies is
\begin{align}\label{eq:frequency}
 E_e^{(2)} &= \frac{|\bra{g}V\ket{e}|^2}{E_e-E_g} +  \frac{|\bra{f}V\ket{e}|^2}{E_e-E_f} = \frac{(2\pi  E_L\delta\varphi)^2}{-\epsilon_2-\alpha\epsilon_1}=- E_f^{(2)},
\end{align}where $\alpha=\epsilon_1/4\pi E_L$. So the second-order derivative of the frequency difference is:
\begin{align*}
\partial_{\Phi}^2 \omega_{ef} = \partial_{\Phi}^2 (E_f^{(2)}-E_e^{(2)}) = \frac{16\pi^2E_L^2}{\epsilon_2+\alpha\epsilon_1}.
\end{align*}

Similarly, for the half-integer fluxonium 
we can obtain $\partial_{\Phi}^2 \omega_{ge} = \partial_{\Phi}^2 (E_e^{(2)}-E_g^{(2)}) = {4\pi^2E_L^2}/{\epsilon_1}$.
The ratio between the second-order derivatives of the two configurations is $(\partial_{\Phi}^2 \omega_{ef})/(\partial_{\Phi}^2 \omega_{ge} ) = 4\epsilon_1/(\epsilon_2+\alpha\epsilon_1)$.

When the two qubits have the same frequency ($\omega_{ge}(\varphi_\text{ext}=\pi)=\epsilon_1 , \omega_{ef}(\varphi_\text{ext}=0)= \epsilon_2+\alpha\epsilon_1$, which comes from diagonalizing Eq. \ref{hamiltonian:int} and the double well tunneling hamiltonian of the half-integer fluxonium, their second-order derivatives differ by a factor of 4. This indicates that the integer fluxonium remains similarly first-order insensitive to flux noise in standard frequency regimes.

\begin{figure}
    \centering
    \includegraphics[width=1\linewidth]{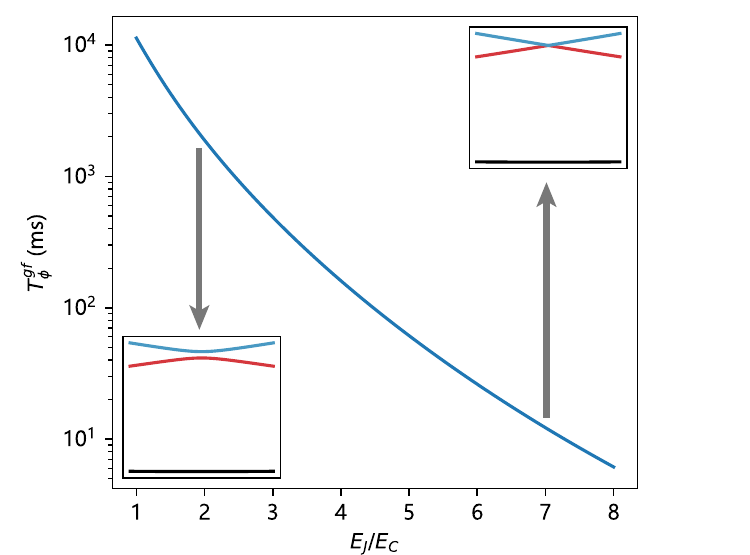}
    \caption{A bigger capacitance results in sharper avoided crossing, which suppresses $\ket{f}\rightarrow\ket{e}$ decay but increases flux noise sensitivity level of frequency, decreasing $T_\phi^{gf}$. The numerical values are obtained with the parameters of the example g-f qubit.}
    \label{fig:curvature}
\end{figure}

Fig.~\ref{fig:curvature} highlights the trade-off between the avoided crossing's sharpness and flux noise sensitivity of the IFQ $\ket{e}-\ket{f}$ subspace. A larger capacitance increases the sharpness, which suppresses \(\ket{f} \rightarrow \ket{e}\) decay but raises the sensitivity to flux noise and lower $T_\phi$ within the g-e-f manifold. 

\textbf{Matrix element:} For the e-f qubit, the matrix elements of the phase and charge operators provide insight into the interaction strengths between the computational states. After diagonalizing the integer fluxonium Hamiltonian (Eq. \ref{hamiltonian:int}) and obtaining the eigenvectors \(\ket{e}\) and \(\ket{f}\) in the local well basis, the phase matrix element is calculated as follows:

\begin{align*}
\bra{e}\hat{\varphi}\ket{f} & \approx 
\frac{1}{2}\begin{pmatrix}
1&0&-1
\end{pmatrix}\begin{pmatrix}
-2\pi& & \\
 & 0 & \\
 & & 2\pi\\ 
\end{pmatrix}\begin{pmatrix}
1 \\
2\alpha\\
1 
\end{pmatrix}=2\pi.
\end{align*}
which turns out to be twice that of the half flux quantum fluxonium. From the phase matrix element, we can calculate the charge matrix element between $\ket{e}$ and $\ket{f}$: 
\begin{align*}
\bra{e}\hat{n}\ket{f} \approx \frac{E_e-E_f}{8E_C}\bra{e}\hat{\varphi}\ket{f} = \frac{2\pi}{8E_C}(\alpha\epsilon_1 + \epsilon_2).
\end{align*}

\begin{figure}
    \centering
    \includegraphics[width=0.7\linewidth]{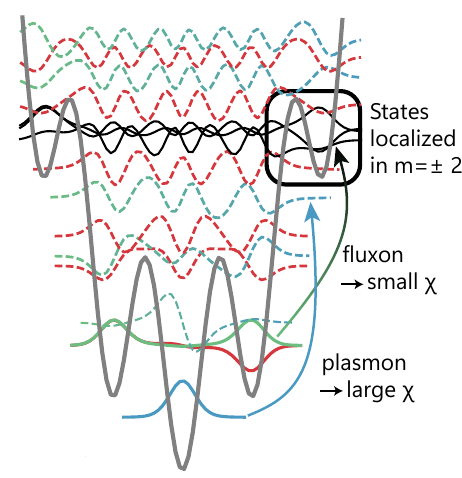}
    \caption{The dispersive shift from a fluxonium state is affected by all transitions from that state. At high resonator frequency, the dispersive shift $\chi$ is mostly affected by the closest detuned fluxonium transition. When the transition is fluxon- or parity-forbidden, $\chi$ is small; when the transition is plasmon, $\chi$ is large. In particular, the transitions that excite the g-e-f manifold to states localized in $m\pm 2$ potential wells are fluxons. The difference between fluxon and plasmon transitions offers a partial explanation to the $\chi$ matching of the example e-f qubit in this paper.}
    \label{fig:fluxon_plasmon}
\end{figure}

\section{Coherence time estimates} \label{sec:coherence}

In this section, we use simple noise models to estimate the lifetimes of the example e-f and the g-f qubits considered in the paper, based on their frequencies and matrix elements.

\subsection{Noise model}\label{sec:noise_model}

We consider two dominant contributions to fluxonium decoherence: dielectric loss and $1/f$ flux noise. These describe the noise model in both the usual frequency range and the low-frequency regime.

The primary source of error in superconducting qubits is energy exchange between the capacitor electric field and microscopic two-level system defects. This is commonly referred to as dielectric loss. The relaxation or excitation rate for a transition $j \rightarrow i$ with frequency $\omega_{ij} = \omega_{j}-\omega_{i}$ can be expressed as \cite{Sun2023characterization,zhang2021universal,SmithThesis,nguyen2019highcoherence}:

\begin{align}\label{eq:dielectric}
\Gamma_{1,diel}^{ij} &= \frac{16 E_C}{ \hbar Q_{\text{cap}}}  |\bra{i}\hat{n}\ket{j}|^2\times \frac{1}{2} |N_P(-\omega_{ij},T)|,
\end{align}
where the Planck's function term reflects the interaction strength between qubit and environment as a function of environment temperature $T$:
\begin{align*}
|N_P(-\omega_{ij},T)|= |\frac{1}{e^{-\hbar \omega_{ij}/k_B T}-1}|=|\text{coth}(\frac{\hbar \omega_{ij}}{2k_B T}) + 1|.
\end{align*}
We assume a frequency-independent capacitance quality factor $Q_{\text{cap}} = 1/\tan \delta_C = 10^{5}$, because underlying mechanisms such as two-level system density do not typically exhibit strong frequency dependence in the range of interest. We use a typical dilution refrigerator temperature of $T = 20$ mK.

In addition to dielectric loss, $1/f$ flux noise could be a significant contributor to dissipation in the very low-frequency regime. The bit-flip rate that $1/f$ flux noise causes is proportional to the square of noise amplitude $A_{1/f}$ and inversely proportional to $\omega^{\eta}$, with $\eta\approx1$ \cite{zhang2021universal,Sun2023characterization,yan2016flux}.
For simplicity, we use $A_{1/f} = 1\mu \Phi_0$ with:
\begin{align}
\Gamma_{1,1/f}^{ij} &= |\bra{i}\hat{\varphi}\ket{j}|^2 (\frac{E_L}{\hbar \Phi_0} )^2 \frac{2\pi A_{1/f}^2 }{|\omega_{ij}|}
\end{align}

The combined dissipation rate is then given by:
\begin{align}\label{eq:T1}
\Gamma_1^{ij} = \Gamma_{1,diel}^{ij} + \Gamma_{1,1/f}^{ij}.
\end{align}

Noise channels contributing to dephasing include both phase smearing due to variations in the qubit frequency induced by external flux fluctuations and photon-shot noise from resonator photons performing weak measurements.  In our qubit-resonator setup, we minimize the difference between the dispersive shifts of the two computational states, thereby reducing thermal photon shot noise. Consequently, we expect most idling dephasing in the e-f and the g-f qubit to arise from second-order coupling between flux noise and qubit frequency. 
We estimate Ramsey dephasing time with\cite{ithier2005decoherence,nguyen2019highcoherence}:
\begin{align}\label{eq:Tphi} 
\Gamma_{\phi} =A_{1/f}\partial_{\lambda}\omega_{ij}+ A_{1/f}^{2} \partial_{\lambda}^{2}\omega_{ij},
\end{align}
and use $1/f$ noise amplitude $A_{1/f}=1$ $\mu \Phi_0$.

\subsection{Balance between dielectric loss and $1/f$ loss for e-f qubit computational subspace lifetime}\label{ef_coherence}
In Appendix \ref{sec:analytical}, we derived that due to greater wavefunction delocalization at the integer sweet spot, the phase matrix element for the e-f qubit is twice that of the HFQ fluxonium: $\bra{e}\hat{\varphi}_{IFQ}\ket{f}=2\bra{g}\hat{\varphi}_{HFQ}\ket{e}.$ Additionally, the frequency curvature of the e-f qubit is four times that of the HFQ fluxonium at the same frequency: $\partial_{\Phi}^2 \omega_{ef}(\varphi_\text{ext}=0) = 4\partial_{\Phi}^2 \omega_{ge}(\varphi_\text{ext}=\pi)$. These two minor differences imply that the e-f subspace decoheres similarly to an HFQ fluxonium with the same frequency. 

\begin{figure}
    \centering
    \includegraphics[width=1\linewidth]{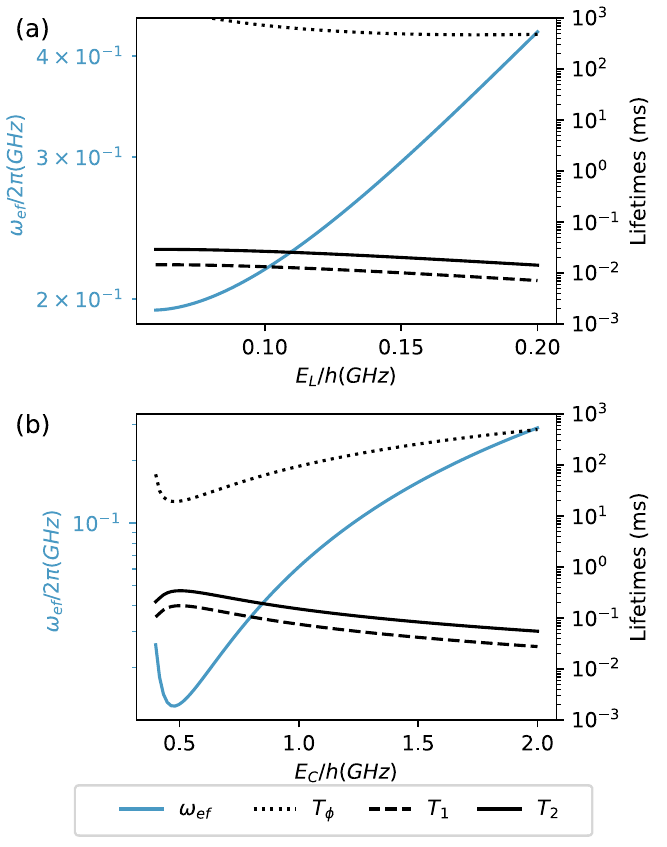}
    \caption{The frequency (blue) and coherence time (black) of the e-f qubit with parameter $E_L$ ($E_C$) modified from the example e-f qubit while keeping $E_J$, $E_C$ ($E_L$) fixed. The coherence times are computed using the noise model in \ref{sec:coherence}. The example e-f qubit parameter is roughly at a local $T_2$ optima.}    
    \label{fig:ef_T1}
\end{figure}

However, the e-f subspace scales toward lower frequencies more rapidly than conventional fluxonium: $\omega_{ef}^{IFQ} \approx \epsilon^2_1/(4\pi^2E_L)$ compared to $\omega_{ge}^{HFQ} = \epsilon_1$, where $\epsilon_1$ is the tunneling amplitude between adjacent potential wells. The frequency scaling from changing $E_L$ or $E_C$ in the example e-f qubit is illustrated by the blue curves in Fig.~\ref{fig:ef_T1} (a)(b).

We also plot the coherence times estimated with noise parameters from Appendix \ref{sec:noise_model} in Fig.~\ref{fig:ef_T1}. Note that when $\omega_{ef}$ drops below $2\pi \times 10 \text{MHz}$, $T_1^{ef}$ gets limited by 1/f flux noise, suggesting that going to ultra-low frequency is not desirable. Additionally, we observe a trade-off between $T_1$ and $T_\phi$. Lower frequencies suppress dielectric loss but decrease dephasing time. The circuit parameters of the example e-f qubit are roughly at the locally optimal $T_2$.          

\subsection{e-f qubit erasure channel}
Due to a relatively high frequency of the g-e transition compared to the environmental temperature, decay from $\ket{e}$ to $\ket{g}$ is one-way. The charge matrix element $\bra{g}\hat{n}\ket{e}$ is $\sqrt{2}$ times that of the conventional fluxonium qubit at half-integer flux quantum \cite{mencia2024integer}. Thus, this dielectric-loss limited transition remains suppressed by small matrix elements. This long lifetime is also explained by parity protection \cite{ardati2024bifluxon}. This decay from $\ket{e}$ to $\ket{g}$ is what we convert to erasure in the e-f qubit.

\subsection{g-f qubit computational subspace error}

\subsubsection{Balance between $T_\phi$ and $T_1^{ef}$ for the g-f qubit}

The e-f subspace frequency $\omega_{ef}\approx \epsilon_1^2/16\pi^4E_L^2$ corresponds to the sharpness of the avoided crossing, where $\epsilon_1$ represents the single-well tunneling amplitude (discussed in Appendix \ref{sec:analytical}). A sharper avoided crossing is associated with a longer lifetime for the erasure transition $\ket{f}\rightarrow\ket{e}$, which reduces the erasure error rate. However, it also increases the sensitivity of the g-f frequency to $1/f$ noise, thereby elevating the computational subspace error rate. For optimal performance in QEC, a balance between the erasure error rate and the computational subspace error rate should be engineered. To illustrate the trade-off, we estimate the coherence times using the equations from Section \ref{sec:noise_model} with experimentally motivated noise parameters $Q_{cap}$ of $10^{5}$ and $A_{1/f}=10^{-6}$. By sweeping across different $E_C$ values while holding $E_J$ and $E_L$ constant, we calculate $T_\phi$ and $T_1^{ef}$, plotting the results against $\omega_{ef}$ in Fig.~\ref{fig:gf_coherence} (a). As shown in the plot, the erasure ratio, approximated as $T_1^{ef}/(T_1^{ef}+T_\phi^{gf})$, monotonically decreases as $\omega_{ef}$ decreases, consequently, an optimal $\omega_{ef}$ can be found where $T_\phi^{gf}$ is very long, and $T_1^{ef}$ is very (yet not ridiculously) small.

\subsubsection{g-f qubit sensitivity to non-zero $\varphi_\text{ext}$}

A small deviation of the bias flux $\varphi_\text{ext}$ away from zero enables a direct transition from $\ket{f}$ to $\ket{g}$. For the e-f qubit, non-zero $\varphi_\text{ext}$ slightly un-biases the erasure error and reduces the computational subspace lifetime, which is relatively manageable. However, for the g-f qubit, non-zero $\varphi_\text{ext}$ can cause direct computational bit-flip errors, which is significantly worse if it spoils the erasure-dominant error structure. Additionally, the e-f qubit frequency loses its first-order insensitivity to flux fluctuations, introducing a first-order term, $2A_{1/f}^2 (\partial_{\Phi} \omega_{ij})^2 |\ln(\omega_{\text{ir}} t)|$ into the curly braces of Eq. \ref{eq:Tphi}, which again could reduce the dephasing time and be damaging \cite{groszkowski2018coherence}.

\begin{figure}
    \centering
    \label{fig:gf_coherence}
    \includegraphics[width=1\linewidth]{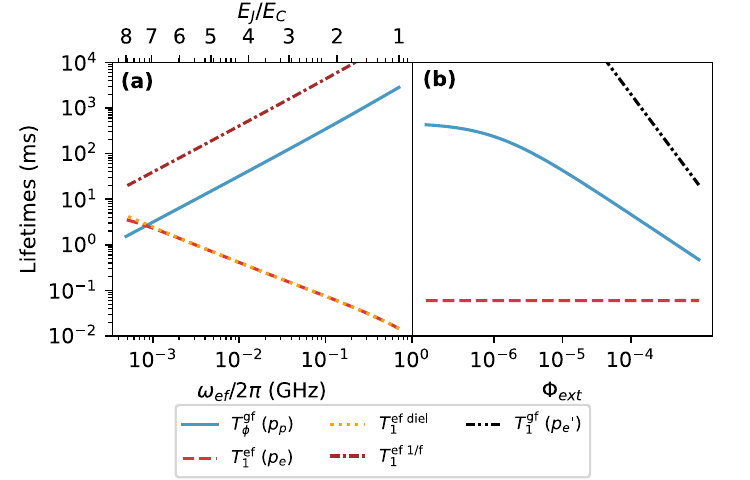}
    \caption{\textbf{(a)} Computational subspace dephasing time ($T_\phi^\text{gf}$, corresponding to computational subspace error rate) and leakage lifetime ($T_1^\text{ef}$ corresponding to erasure error rate) for the g-f qubit, as a function of e-f transition frequency by varying $E_C$ of the example g-f qubit. Coherence times are computed with the noise model in Section \ref{sec:noise_model}. A desirable error hierarchy (erasure ratio) can be achieved at a specific $E_J/E_C$ ratio, depending on noise parameters, such that the sweet spot is sweet, and erasure error does not overwhelm erasure conversion. \textbf{(b)} Sensitivity of $T_\phi^{gf}$, and the decay lifetime ($T_1^{gf}$) of $\ket{f}\rightarrow\ket{g}$ in non-zero external field $\varphi_\text{ext}$ for the example g-f qubit.}
\end{figure}

To quantify the impact of non-zero $\varphi_\text{ext}$ on the g-f qubit, we numerically calculate the \(\ket{f} \to \ket{g}\) decay lifetime $T_1^\text{gf}$, the dephasing time $T_\phi^{gf}$, and erasure lifetime $T_1^\text{ef}$.  The results are shown in Fig.~\ref{fig:gf_coherence} (b). Even at a conservative $\varphi_\text{ext} \approx 10^{-5}$, the direct decay lifetime  $T_1^\text{gf}$ remains on the timescale of seconds, making direct decay negligible. The computational dephasing time $T_\phi^{gf}$ is only weakly affected for $\varphi_\text{ext}<10^{-6}$, and the erasure lifetime $T_1^{ef}$ remains largely insensitive to flux bias. These results demonstrate that the g-f qubit is robust against small flux bias drifts, maintaining a low sensitivity to non-zero $\varphi_\text{ext}$.

\subsection{How often detection is needed for the g-f qubit}\label{sec:how_often}

Since both dephasing within the computational subspace and decay from \(\ket{f}\) to \(\ket{e}\) are first-order processes, the rates of phase flips and leakage to \(\ket{e}\) are approximately constant over short timescales. These rates are related by $p_{\phi}/p_{e} = \Gamma_\varphi^{gf} / \Gamma_1^{ef}$. In contrast, cascaded bit-flip errors from \(\ket{f}\) to \(\ket{g}\) via \(\ket{e}\) are second-order processes, with rates depending on the population in \(\ket{e}\). As the population in \(\ket{e}\) increases, cascaded bit-flips accumulate more rapidly, eventually making the ratio of bit-flip errors to leakage errors approach unity. The frequency of leakage detection strongly influences the ratio of computational bit-flip errors to erasure errors. To preserve the erasure-dominant error structure, we need to determine how quickly direct bit-flip errors become significant.

Assuming erasure conversion is instantaneous, $p_\text{bit}/p_{e}$ can be roughly approximated as $\langle g \rangle / \langle e \rangle$ when the qubit starts in \(\ket{f}\). To analyze this, we model the population dynamics of the lowest three energy levels during decay using the following system of equations:

\[
\frac{d}{dt} 
\begin{pmatrix} 
P_0(t) \\ P_1(t) \\ P_2(t) 
\end{pmatrix} 
= 
\begin{pmatrix} 
0 & \Gamma_{10} & 0 \\
0 & -\Gamma_{10} - \Gamma_{12} & \Gamma_{21} \\
0 & \Gamma_{12} & -\Gamma_{21}
\end{pmatrix} 
\begin{pmatrix} 
P_0(t) \\ P_1(t) \\ P_2(t) 
\end{pmatrix},
\]
where $\Gamma_{ij}$ are decay rates estimated using $T = 20$ mK, $Q_{cap}=10^{5}$, and $A_{1/f}=10^{6}$.

\begin{figure}
    \centering
    \label{fig:gf_idling_dynamics}
    \includegraphics[width=0.8\linewidth]{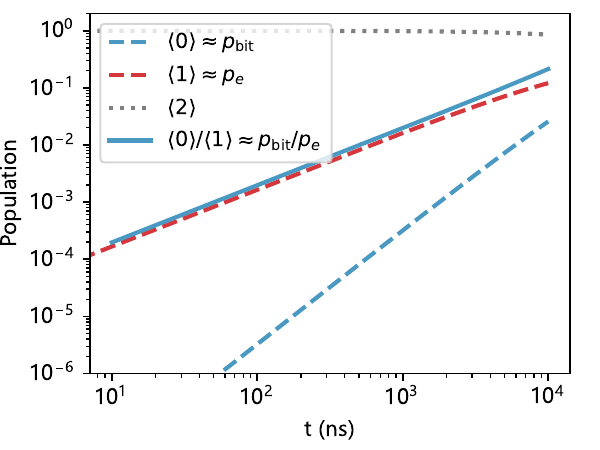}
    \caption{Idling population dynamics when initialized in $\ket{2}$. The populations of the intermediate state \(\ket{1}\) (dashed red line) and the ground state \(\ket{0}\) (dashed blue line) follow a power-law dependence on idling time t. The solid blue curve represents the ratio $\langle 0 \rangle / \langle 1 \rangle$, which reflects the contributions of cascaded bit-flip errors relative to erasure errors. This ratio determines how often erasure conversion should be performed to prevent excessive undetectable errors.}
\end{figure}

We simulated the decay dynamics using QuTip, initializing the qubit in \(\ket{2}\) and showed the population dynamics of the three qubit levels over time in Fig.~\ref{fig:gf_idling_dynamics}. The populations of the intermediate state \(\ket{1}\) (dashed red line), the ground state \(\ket{0}\) (dashed blue line) as well as the ratio $ \langle 0 \rangle / \langle 1 \rangle$ (solid blue line) increase following a power-law trend with respect to idling time t. This ratio $ \langle 0 \rangle / \langle 1 \rangle$ quantifies the amount of cascaded bit-flip errors to erasure errors and determines the necessary frequency of erasure conversion to maintain an erasure-dominant error structure. Under the chosen noise model, performing erasure conversion every 1 millisecond is sufficient to achieve a high erasure ratio and to limit the impact of undetectable errors.

\subsection{Decay rate and branching ratio for the g-f qubit considered}
For the noise model mentioned in \ref{sec:noise_model}, we list the lifetime and branching ratio for the first 7 levels of the example g-f qubit considered in this paper in Table \ref{tab:branching_ratio}.

\begin{table}
\caption{\label{tab:branching_ratio}%
Lifetimes ($\mu s$) of the example g-f qubit considered in this paper. The decay or heating rates are calculated with the noise model in \ref{sec:noise_model}.}
\begin{ruledtabular}
\begin{tabular}{ccccccccccc}
\multicolumn{2}{c}{ }&\multicolumn{7}{c}{Branching Ratio} \\\cmidrule{3-9}%
 &lifetime &0&1&2&3&4&5&6\\
\hline
0&8764& &1.00& & & & & \\

1&20.12&0.76& &0.24& & & & \\

2&59.40& &0.99& & & & & \\

3&2.15&0.50& &0.49& & & & \\

4&1.64& &0.35& &0.63& &0.03& \\

5&1.95&0.06& &0.20& &0.65& &0.09\\

6&3.18& & & &0.03& &0.77& \\
\end{tabular}
\end{ruledtabular}
\end{table}

\subsection{Outlook for coherence time improvements of e-f and g-f qubit in terms of noise rates}
We note that the difficulty in suppressing $1/f$ noise through advancements in fabrication techniques and materials could pose a challenge to the future development of the e-f qubit. The improvements in $T_1^{ef}$ by suppressing $\omega_{ef}$ are limited by $1/f$ flux noise. However, compared to suppressing $1/f$ flux noise, the industry is spending more effort on minimizing dielectric losses by reducing the density of two-level system defects \cite{mohseni2024build}. This trend benefits the future potential of qubits limited by dielectric loss, such as the transmon qubit or the g-f qubit in this paper. Similarly, a higher quality factor of the capacitance means the g-f qubit can have a larger e-f splitting, which enhances its dephasing time and, therefore, its effective coherence time. 

\section{Single qubit gate simulations}

\subsection{Fidelity measure}\label{fidelity_measure}

We model dissipation via the Lindblad master equation, incorporating the set of collapse operators $\{L_k\}$ defined as:
\begin{equation}\label{decay_terms}
    \left\{ \sqrt{\Gamma_{1}^{ij}} |j\rangle\langle i| \right\}_{i \neq j} \cup \left\{ \sum_{k} \sqrt{\Gamma_{\varphi}^{k}} |k\rangle\langle k| \right\}
\end{equation}
where the relaxation rates $\Gamma_1^{ij}$ and dephasing rates $\Gamma_\varphi^{i}$ are derived from the noise model detailed in Section \ref{sec:noise_model}. To quantify performance, we calculate the computational subspace infidelity by averaging the squared fidelity over the six cardinal states of the Bloch sphere \cite{bowdrey2002fidelity}:

\begin{widetext}
\begin{align}\label{fidelity}
    1-\mathcal{F} = 1- \max_{\theta} \left\{\frac{1}{6}\sum_{\rho_0\in \{\pm\ket{X}, \pm\ket{Z},\pm\ket{Y}\}} \Bigg(\text{Tr}(\sqrt{\rho_\text{ideal}} \rho_\text{actual}(\theta)\sqrt{\rho_\text{ideal}} )\Bigg)^2\right\} 
\end{align}
\end{widetext}
where:
\begin{align}\label{fidelity_second}
    \rho_\text{actual}(\theta) = R_z(\theta)\frac{\mathcal{G}\rho_0^{(2\times2)}}{\text{Tr}(\mathcal{G}\rho_0^{(2\times2)})}R_z(\theta)^\dagger,
\end{align}
$\mathcal{G}\rho_0^{(2\times2)}$ describes the truncated 2 level density matrix at the end of the pulse, $\rho_\text{ideal} = X \rho_0 X^\dagger$ is the ideal final state (suppose we are studying an X gate), the trace in the denominator is for renormalization, and $R_z(\theta)$ compensates for a phase factor via virtual  Z -rotation.

\subsection{the g-f qubit Raman gate simulation details}\label{raman_details}

We numerically simulate a capacitively driven X gate using the $\ket{g}\leftrightarrow \ket{h} \leftrightarrow \ket{f}$  $\Lambda$ structure as shown in Fig.~\ref{fig:gates}. The gate involves two simultaneous $\sin^2$-shaped drives at frequencies $\omega_{gh}-\Delta$ and $\omega_{fh}-\Delta$ with amplitudes $A_{03}$ and $A_{23}$. The amplitude ratio $A_{gh} / A_{fh}$ is inversely proportional to the charge matrix element ratio $\bra{g}\hat{n}\ket{h}/\bra{f}\hat{n}\ket{h}$. 

To optimize the gate, we explore combinations of detuning $\Delta$ and gate duration $T_\text{gate}$. A gradient-free optimization routine adjusts the drive amplitudes to maximize the target population at the end of the pulse. After optimization convergence, we simulate the gate with dissipation modeled by collapse operators from \eqref{decay_terms}
using the method, temperature, and noise amplitudes described in Appendix \ref{sec:noise_model} to estimate gate performance under realistic noise conditions.

We report the computational subspace error rate $1-\mathcal{F}_\text{cond.}$, probability $p_e$ of leakage to $\ket{e}$, and  probability $p_e'$ of leakage to states outside the g-e-f manifold, averaged over initial states $\{\pm\ket{X}, \pm \ket{Y},\pm \ket{Z}\}$, in FIG.~\ref{fig:gates} (d).

As shown in Fig.~\ref{fig:gates}, increasing the detuning in the first $0.4$ GHz reduces the computational subspace error by suppressing decay from the intermediate state back to the computational subspace. The ratio of leakage to $\ket{e}$ relative to computational subspace errors in good detuning regions indicates effective preservation of an erasure-dominant error structure. This result indicates that gates for the g-f qubit should utilize cleverly chosen detuning to minimize direct bit-flip caused by intermediate state decay. 

The large drive amplitudes required by large detuning induce far-off-resonant excitations to other dipole-allowed transitions (e.g. $\ket{5}$ and $\ket{7}$). While their populations return to minimal levels at the end of the pulse, these states can decay into ``intermediate states" that couple to $\ket{4}$ and $\ket{6}$, introducing additional leakage. Since this leakage is comparable in magnitude to computational subspace errors, it can lead to correlated errors unless mitigated.

\begin{figure}
    \centering
    \includegraphics[width=1\linewidth]{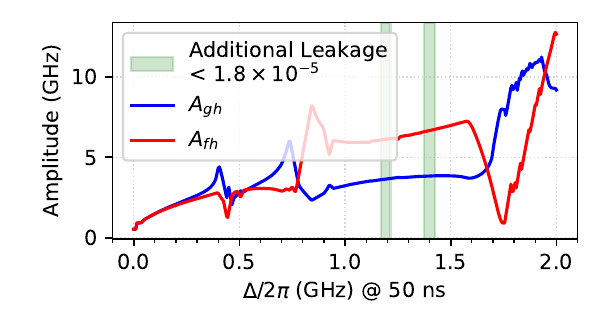}
    \caption{Drive amplitude on the detuned $\ket{g}-\ket{h}$ and $\ket{f}-\ket{h}$ transition in the optimized  single qubit Raman gate. Admittedly, the drive amplitude is quite high in regions with low leakage error (green shaded regions).}
    \label{fig:raman_amplitude}
\end{figure}

We acknowledge that far-detuned Raman processes will incur a power-requirement price tag. In Fig.~\ref{fig:raman_amplitude}, we show the drive amplitudes for the optimized gate.

\section{Details of numerically simulating the leakage detection readout}
\subsection{Numerical Simulation Methods}\label{sec:detection_numerical_details}

To simulate the dispersive readout process, we perform Monte-Carlo simulations\cite{PhysRevLett.68.580} using the Hamiltonian Eq. \eqref{hamiltonian} in QuTip\cite{qutip1},  with initial states that are the dressed fluxonium eigenstates coupled to the resonator vacuum state: $y_0 = \{\ket{a}, \ket{b}, \ket{L}\} \otimes \ket{0}_r$, where the resonator decay term is $\sqrt{\kappa} \hat{a}$ and the decay rate is $\kappa$, and a drive term:
\[
\mathcal{H}_{\text{drive}} = A(t)\cos((\omega_r+\chi_g) t)\hat{n}_r
\]
where $A(t)$ is the pulse amplitude function. Note that in the readout simulations presented in this paper, we do not include qubit decay terms, since we only aim to understand the effect of the dispersive erasure detection via the resonator. For simplicity in finding the stop time of the pulse through simulation, we use a square pulse envelope of amplitude $A$ with a fixed 20 ns long sin$^2$ smooth ramping up, and report the evolution of states (density matrices) truncated at the first timestep when photon number\footnote{Photon number is associated with the coherent state $\alpha$ by $n_p = |\alpha|^2$.} associated with computational states returns to a local minimum as described in \ref{subsec:dispersive_detection}.

\subsection{Amplitude damping analysis}

We expect any amplitude damping in the computational subspace to be minimal because the qubit's transition between computational states should be either very far-detuned from the resonator frequency (the e-f qubit) or forbidden in the first place. However, the simulation reveals a bit-flip error rate on the order of $10^{-5}$ in the example g-f qubit when decay on resonator is turned on. This resonator decay-dependent bit-flip is consistent with the Purcell effect.

Upon checking the dressed ``cross matrix elements" (in the language of cross-resonance two qubit gates), we found $|\bra{e_f n_r}\hat{a}_n\ket{f_f n_r}|\approx 4\times 10^{-6}$ for the e-f qubit coupled to resonator level $n$, and $|\bra{g_f n_r}\hat{a}_n\ket{e_f n_r}|\approx 2.5\times 10^{-3},|\bra{e_f n_r}\hat{a}_n\ket{f_f n_r}|\approx 7\times 10^{-4}$ for the g-f qubit, which suggest Purcell enhanced fluxonium decay may contribute to what we saw in numerical simulation in addition to the dephasing errors. Another qubit error source might be the partial trace operation we did numerically to isolate the qubit density matrix.

\begin{figure}
    \centering
    \includegraphics[width=0.7\linewidth]{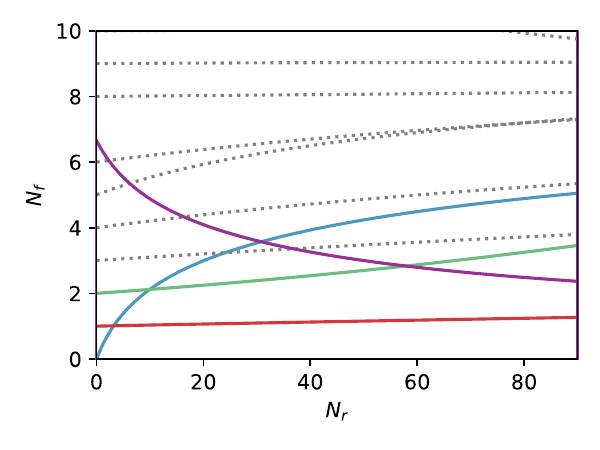}
    \caption{In the example e-f qubit readout simulation, the hybridization of the $\ket{g}_q$ branch and $\ket{7}_q$ branch is slow, which prevents rapid ionization of fluxonium population during readout.}
    \label{fig:hybridization}
\end{figure}

In addition to transitions from the computational states, the readout drive can also cause secondary leakage of the leaked state. When a fluxonium transition is near the resonator frequency, it hybridizes with the resonator transition. If the hybridization is too rapid, the qubit population could leak, like transmon ionization \cite{ionization,ionization_new}. To understand the structure of hybridization, we use the method in \cite{ionization} to label dressed states by product-state indices. 
We found the hybridization to be smooth and slow. For the example g-f qubit, we do not observe any crossover of branches. For the example e-f qubit, we plot the crossover of the $\ket{g}$ branch and the $\ket{7}$ branch in \ref{fig:hybridization}. Due to the computational cost and complexity of studying evolution in a large Hilbert space, we leave the question of numerically estimating fluxonium leakage during measurement to future work. However, secondary leakage events occur with a two-step probability $P(\text{leak to }\ket{L})\times P(\text{leak to }\ket{L'})$ where $L'$ represents some other higher state that the original leakage state is coupled to and also near-resonant with the resonator. This suggests that secondary leakage may be less damaging.

\subsection{Quadrature projection of the Q function}
\begin{figure}[h]
    \centering
    \includegraphics[width=1\linewidth]{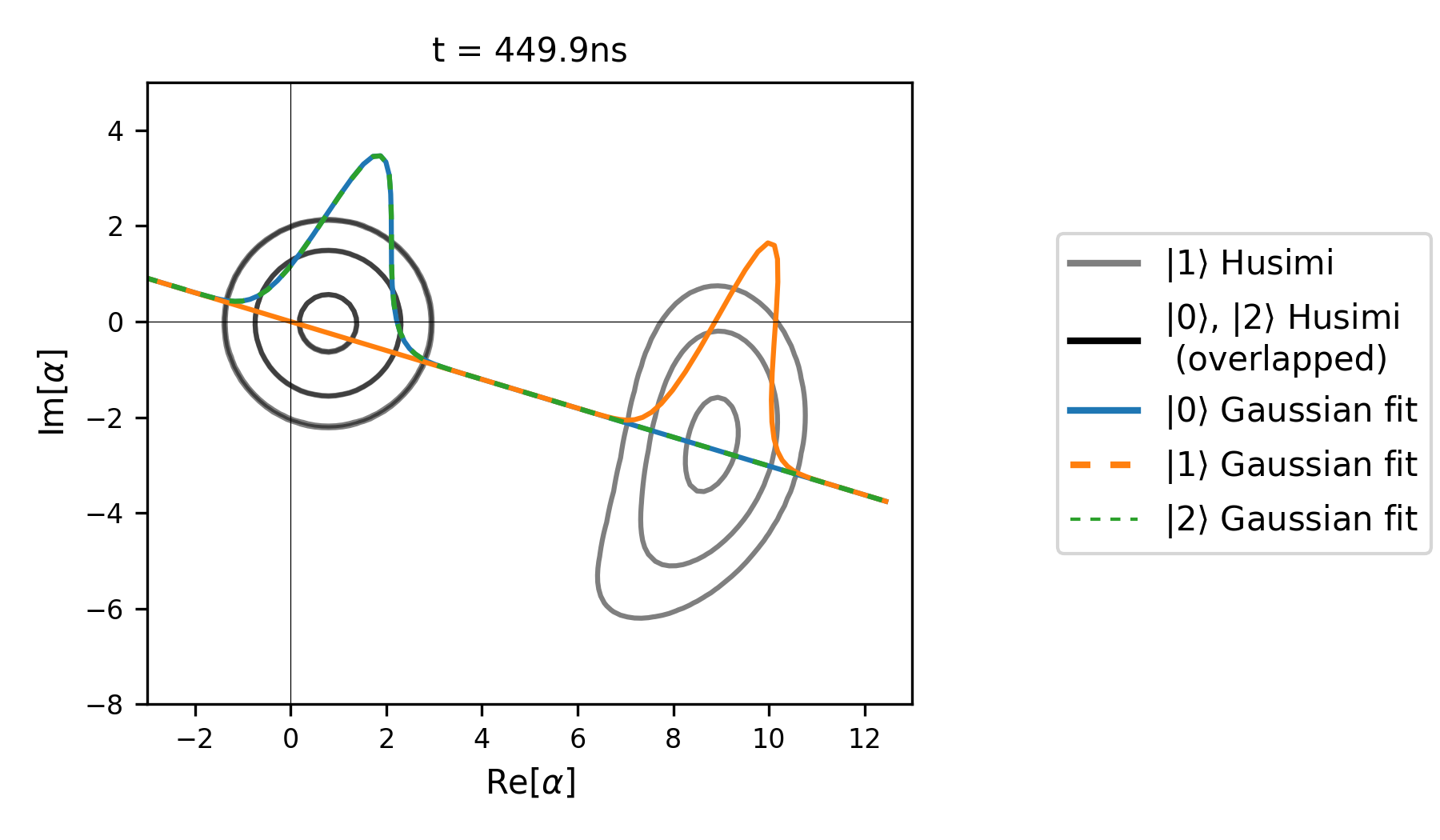}
    \caption{The Husimi-Q function at the last time instance in the simulation of the erasure readout for the g-f qubit. The contours encircle ${0.3,0.9,0.99}$ of the cumulative probability density. From this separation in the projected Husimi-Q function, one can roughly see they are separated, and this is related to SNR.}
    \label{fig:Husimi-snapshot}
\end{figure}

\begin{figure}[h]
        \centering
        \includegraphics[width=1\linewidth]{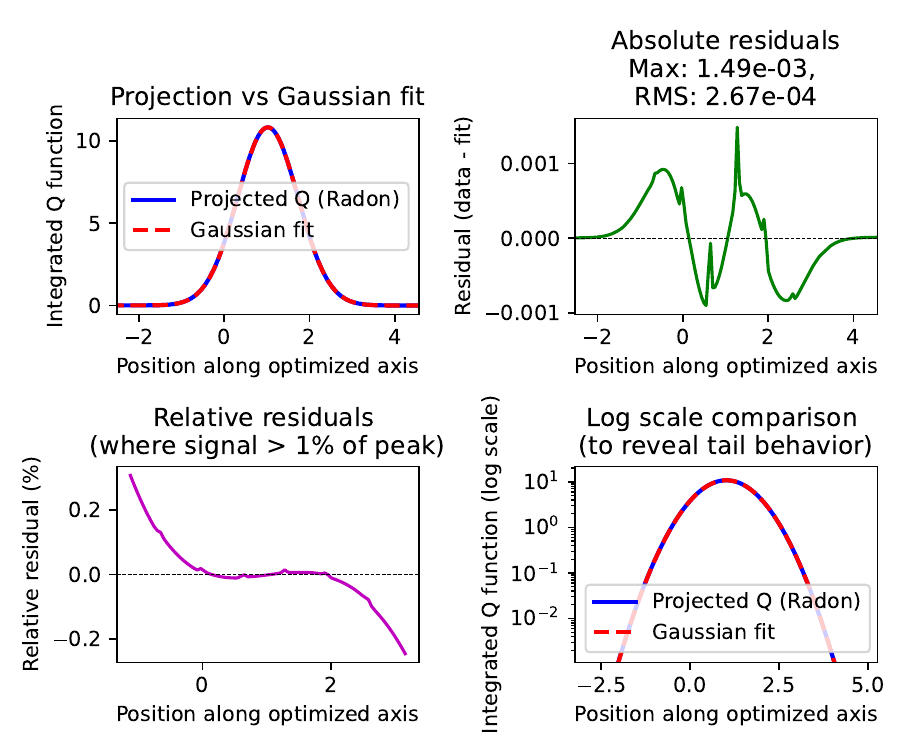}
        \caption{Visualization of how well the projected Husimi-Q function density distribution can be represented as a Gaussian.}
        \label{fig:Husimi-diagnostics}
\end{figure}

In the main text, we used 1D Gaussians to represent the projected Husimi-Q functions. In principle, nonlinear effects in readout can cause the Q function to squeeze and become banana-shaped. Here, we present diagnostics of how well the Gaussian can represent the projected Husimi-Q functions. After we pick the projection axis (optimizing the angle in radon transform to maximize the separation between projected 1d gaussian centers) as illustrated in Fig.~\ref{fig:Husimi-snapshot}, we show the fit and analytics in Fig.~\ref{fig:Husimi-diagnostics}, where the absolute residuals are very small. Although the right and left ``tails" of the projection deviate a bit from Gaussian in a relative sense, the ``bulk" can be described by a Gaussian very well.

\subsection{Other practical concerns}\label{subsec:practical}

Due to the computational cost of simulating large-dimension quantum objects, the readout simulation for each initial state in the e-f and the g-f qubit was done with Monte Carlo simulation with 1000 trajectories in QuTip \cite{qutip1}. To reduce the cost, we truncate the qubit degrees of freedom to only two (when the initial state is in the computational subspace), or one (when the initial state is the leakage state), level(s). A more detailed investigation into the dispersive readout is deferred until future work. Future studies could utilize dense matrix multiplication acceleration techniques, as in \cite{guilmin2024dynamiqs,ionization}.

In practice, further optimization of the protocol should improve readout accuracy. Also, since imperfect erasure conversion is still better than non-heraldable noise, an accuracy with realistic readout efficiencies might be enough in the near term. To improve the overall performance of the QEC protocol, one may further adjust the balance between false positives and false negatives, or use a potential one-way pulse to bring the leaked qubit back to the computational subspace \cite{levine2024dualrail,zuk2024robust}.

\section{Dispersive readout dephasing}\label{sec:analytical_dephasing}

\subsection{Incoherent dephasing}\label{sec:dephasing_incoherent}

The photon-shot dephasing rate, also referred to as the measurement rate, as it describes the dephasing due to loss of information to the outgoing field, is commonly defined as \cite{SmithThesis}:
\begin{align}\label{eq:dephasing}
    \Gamma_\varphi = A_n n  \frac{\kappa\chi^2}{\chi^2 + \kappa^2}.
\end{align}
It describes the traditional readout, in which the drive frequency is in the middle of the resonator frequencies dressed by two computational states. Still, it does not apply to the leakage detection in this work.

Let us derive a suitable formula using a model in which $\omega_r$ is the bare resonator frequency, $\chi_a$ and $\chi_b$ are the dispersive shifts from the two computational states, and $\Delta = -\chi_L$ is the detuning between the drive frequency and the bare resonator frequency. The resonator is driven with amplitude $A$. We use the method from \cite{SmithThesis} with slight modifications; the result is also consistent in appearance with that in \cite{Aashish_dephasing}. The Hamiltonian coupled to the two-level qubit subspace is 
\begin{equation}
    \begin{aligned}
    H(t) = \hbar \omega_r a^\dagger a + \hbar \omega_q \ket{b}\bra{b} 
    + \hbar \chi_b  a^\dagger a \ket{b}\bra{b} \\
    + \hbar \chi_a a^\dagger a \ket{a}\bra{a} 
    +\hbar A \cos(\omega_d t) (a+a^\dagger),
    \end{aligned}
\end{equation}
with $U = e^{i \omega_d a^\dagger a t}$ bringing into rotating frame,
\begin{equation}
    \begin{aligned}
   H' =&  U H(t) U^\dagger - i \hbar U \dot{U}^\dagger \\
   \approx&\hbar \Delta a^\dagger a + \hbar \omega_q \ket{b}\bra{b} 
    + \hbar \chi_b  a^\dagger a \ket{b}\bra{b}  \\
    &+ \hbar \chi_a a^\dagger a \ket{a}\bra{a} 
    + \hbar \frac{A}{2} (a + a^\dagger),
    \end{aligned}
\end{equation}
where $\Delta_{dr} = \omega_r - \omega_d$. In the Lindblad equation that includes resonator decay,
\[
\dot{\rho} = -\frac{i}{\hbar} [H', \rho] + \kappa \left( a \rho a^\dagger - \frac{1}{2} \left( a^\dagger a \rho + \rho a^\dagger a \right) \right),
\]
the off-diagonal qubit component is 
\begin{equation}
    \begin{aligned}
\dot{\rho}_{ab}  &=( -i\frac{A}{2}\alpha_a + i\omega_q +i\frac{A}{2}\alpha_b^* + \kappa\alpha_a \alpha_b^*)\rho_{ab} \\ 
&\quad - (i\Delta \alpha_a  + i \chi_a \alpha_a  +i\frac{A}{2} + \frac{\kappa}{2}\alpha_a )a^\dagger \rho_{ab} \\
&\quad+(i\Delta\alpha_b^*+i\chi_b\alpha_b^* +i\frac{A}{2} -\frac{\kappa}{2}\alpha_b^*)\rho_{ab}a.
    \end{aligned}
\end{equation}
Setting
\[
\alpha_a = -\frac{A}{2(\Delta+\chi_a)-i\kappa}, \alpha_b^* = -\frac{A}{2(\Delta+\chi_b)+i\kappa},
\]
is analogous to reaching steady state, then the exponential decay rate of the off-diagonal element amplitude is equal to 

\begin{equation}
    \begin{aligned}
\Gamma_\varphi^{readout} &= Re[\dot{\rho}_{ab}] \\ &= \frac{2 A^2 \kappa(\chi_a -\chi_b)^2 }{(4(\Delta+\chi_a)^2+\kappa^2)(4(\Delta+\chi_b)^2+\kappa^2)}
    \end{aligned}
\end{equation}

\subsection{Dephasing due to photon fluctuations}\label{sec:dephasing_smearing}

\subsubsection{Static approximation}

Suppose the qubit frequency is linearly dependent on the photon number by the AC-Stark shift amount $\Lambda$:
\begin{equation}
    \begin{aligned}
\omega_q(n) = \omega_0 + n \Lambda.
    \end{aligned}
\end{equation}
When coupled to a coherent state with average photon number $\overline{n}$ (assuming $\overline{n}$ is static), the off-diagonal term, after partial tracing out the resonator, evolves according to 
\begin{equation}
    \begin{aligned}
\langle\rho_{ab}(t)\rangle = \rho_{ab}(0)e^{i\omega_0 t}\langle e^{i\Lambda n t}\rangle.
    \end{aligned}
\end{equation}
By approximating the Poisson distribution $P(n)$ as a Gaussian, and using the Gaussian characteristic function, we get 
\begin{equation}
    \begin{aligned}
\langle e^{i\Lambda n t}\rangle =e^{i\Lambda t \overline{n} } e^{-\frac{1}{2} (\Lambda  t)^2\overline{n}}.
    \end{aligned}
\end{equation}
The real part of the exponent describes dephasing with a Gaussian envelope, when coupled to a static coherent state with average photon number $P(\overline{n})$. 

\subsubsection{Dephasing with time-dependent photon number}

When the resonator photon number distribution is time-dependent, the accumulated qubit phase over an interval \([0, t]\) is

\begin{equation}
\phi = \int_{0}^{t} \Lambda\,n(t') \, dt'.
\end{equation}
We are interested in the real part of $\phi$. We can write
\[
n(t) \;=\; \overline{n}(t) \;+\; \delta n(t),
\]
where \(\overline{n}(t)\) is the (known) time-varying mean $\overline{n}(t) = n_{max}\sin^2(\pi t / t_{max})$ (ignoring resonator decay) and \(\delta n(t)\) is the zero-mean fluctuation. It is the fluctuation that causes phase smearing. The real part in $\phi$ is then the variance 
\begin{equation}\label{coherent_dephasing}
\mathrm{Var}(\phi)
=
\Lambda^2
\int_{0}^{t}
\!\!\int_{0}^{t}
\langle \delta n(t_1)\,\delta n(t_2)\rangle \,dt_1\,dt_2,
\end{equation}  
Using an Ornstein–Uhlenbeck (exponential) Kernel, we write the correlation function as 
\begin{equation}\label{OUkernel}
\langle \delta n(t_1) \,\delta n(t_2)\rangle
\,=\,
\sqrt{\overline{n}(t_1)\,\overline{n}(t_2)} \,
\exp\!\bigl[-\tfrac{|t_1 - t_2|}{\tau_c}\bigr],
\end{equation}
where the correlation time is \(\tau_c = 1/\kappa\), and the variance recovers the Gaussian (\(t^2\)) envelope from the static model. Eqs. \ref{coherent_dephasing}, \ref{OUkernel} is combined in the main text into Eq. \ref{eq:smearing_equation_combined}.

\section{Results on g-f qubit double STIRAP CZ}\label{sec:STIRAP}

\subsection{``safe" intermediate state $\ket{i}$}
\begin{figure}
    \centering
    \includegraphics[width=1\linewidth]{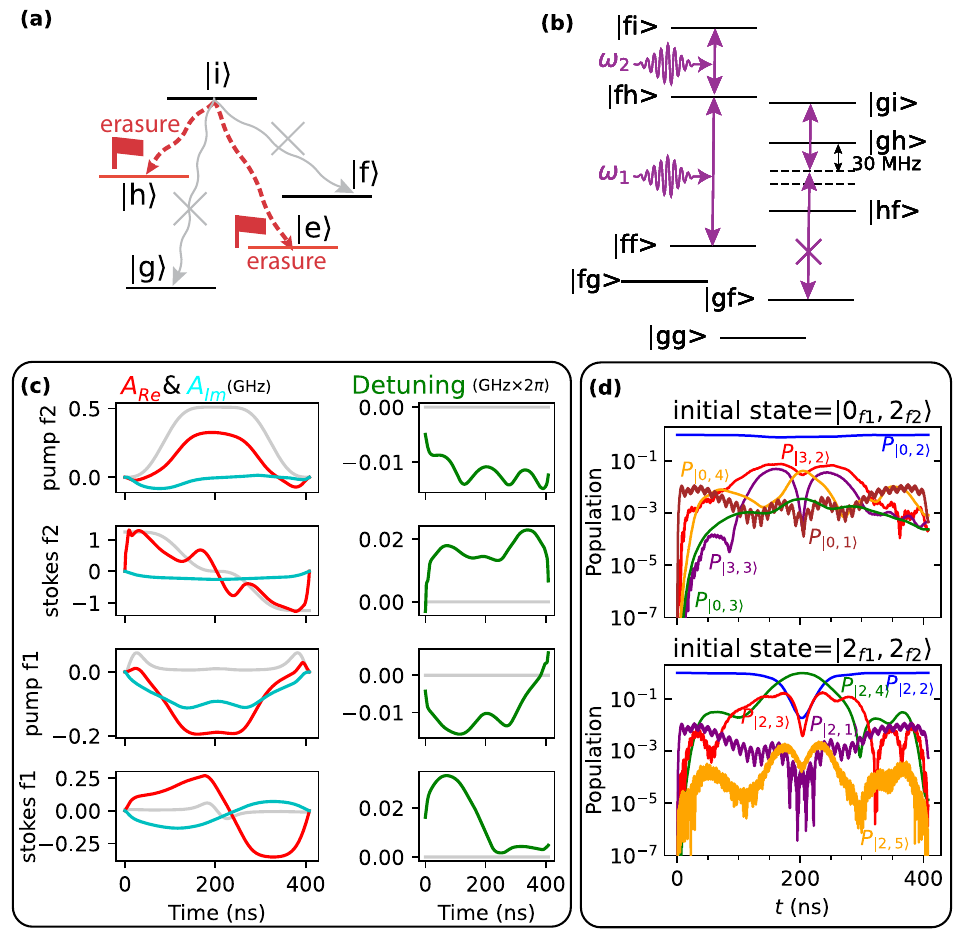}
    \caption{\textbf{(a)} The level diagram that shows that the fourth excited state $\ket{i}$ is a safe intermediate state in geometric CZ gate. The reason is that because state $\ket{i}$ doesn't directly decay to the two computational states by parity selection, if we can detect $\ket{h}$, then we can have a very small error rate in the computational subspace. \textbf{(b)} To use state $\ket{i}$ as the intermediate state in geometric CZ, we need to use STIRAP to reduce exposure to dangerous state $\ket{h}$. STIRAP consists of two drive pulses on two transitions. To disable the unwanted $\ket{gf}-\ket{gh}-\ket{gi}$ transition, we use selective darkening. \textbf{(c)} The numerically optimized pulse that corrects for the nonlinear effects in large drives. The real ($A_{Re}$), imaginary ($A_{Im}$), and time-dependent detuning envelopes for the stokes and pump pulses on the two qubits are shown in red, cyan, and green, respectively. The faint Grey curve in the first column is the initial guess envelope.    \textbf{(d)} Population evolution when simulated without dissipative terms that shows an ideal double-STIRAP process in the wanted $\ket{ff}-\ket{fh}-\ket{fi}$ process, and disabled unwanted $\ket{gf}-\ket{gh}-\ket{gi}$ process. }
    \label{fig:STIRAP}
\end{figure}

Due to parity selection, state $\ket{i}$ (the 4th excited state) can not directly decay to the two computational states, but only decay to state $\ket{e}$, which causes erasure error, or to $\ket{h}$, which might be detected if we add a secondary erasure detection step in addition to detecting $\ket{e}$ (otherwise it still causes undetectable error). Although we do not explicitly show that it is doable in this paper, there are no obvious obstacles to detecting a second erasure state. We show this error structure within a single qubit in Fig.~\ref{fig:STIRAP} (a)).

STIRAP has been used for the CZ gate to minimize decay of intermediate states in Rydberg-atom quantum computing \cite{PhysRevA.101.062309}. In STIRAP, the population is coherently transferred between the initial and final states by following a dark state of the three-level system, decoupled from the intermediate state, when the system is driven adiabatically. This perfectly suits our need to minimize exposure to the state $\ket{h}$.

We use $\ket{fi}$ as the SITRAP terminal state (CZ intermediate state). We pick qubit 1, whose parameters are listed in table \ref{tab:qubit_parameters}, and qubit 2 with $E_J/h=5.65$ GHz, $E_C/h=4.2$ GHz, $E_L=0.12$ GHz. In this configuration, bare $\ket{gh}$ is only $41$ Mhz detuned from bare $\ket{hf}$, causing a $30$ MHz shift on $\ket{gh}$ (Fig.~\ref{fig:STIRAP} (b)). This weak hybridization again makes the conditionality of the Controlled Z gate difficult. We circumvent this by numerically optimizing a smooth-shaped pulse ansatz after first fitting to a $400$ ns superadiabatic transitionless driving amplitude \cite{SATD} envelope shown as faint grey curves in Fig.~\ref{fig:STIRAP} (c), then numerically optimizing the pulse ansatz, turning on time-dependent detuning corrections. A coherent fidelity above 0.999 is achieved at $400$ ns, and we show the optimized complex pulse amplitudes in Fig.~\ref{fig:STIRAP} (c) and the population evolution of the wanted and unwanted transitions in Fig.~\ref{fig:STIRAP}(d).

Future studies should focus on designing a more favorable eigenstate structure that yields cleaner population-evolution trajectories, enabling faster gates. Alternatively, it can benefit from using a coupler \cite{Ding2023FTF,setiawan2023STIRAP}, in either case, the idea of using a safe STIRAP terminal state should still be applicable.

\subsection{Numerical Implementation of Time-Dependent Detuning}
To efficiently simulate time-dependent frequency detuning $\delta(t)$ within the ordinary differential equation (ODE) solver, we avoid dynamically integrating frequency shifts during the simulation steps. Instead, the instantaneous detuning is incorporated as a pre-computed cumulative phase correction $\theta_{\text{det}}(t)$, defined as: 
$$\theta_{\text{det}}(t) = \int_{0}^{t} \delta(t') dt'$$
In our numerical implementation, this integral is pre-evaluated over a discrete time grid and dynamically interpolated at each solver step. Given the instantaneous amplitude $A(t) = A_\text{Re}(t) + iA_\text{Im}(t)$, the rotation is applied directly to the carrier argument:
\begin{equation}
    \epsilon(t) = \text{Re} \left[ A(t) \exp\left(i \left[ 2\pi \omega_d t + \theta_{\text{det}}(t) \right]\right) \right]
\end{equation}
where $\omega_d$ is the base carrier frequency, and $\epsilon(t)$ is the coefficient before the driven operator in the lab frame Hamiltonian.

\subsection{Auto-differentiated pulse optimization}
\begin{figure}
    \centering
    \includegraphics[width=1\linewidth]{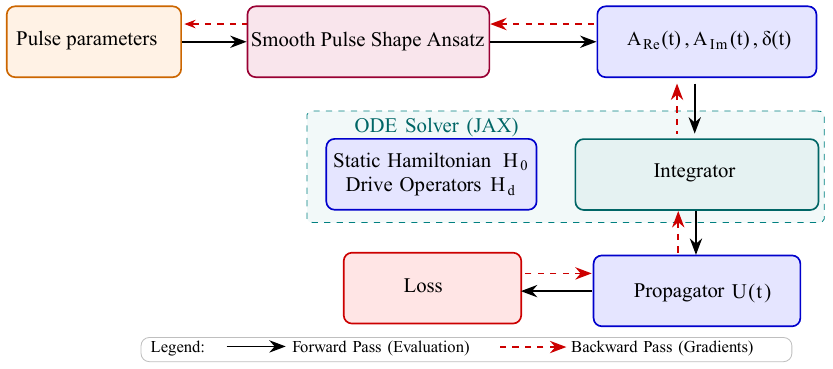}
    \caption{Diagram showing the process of auto-differentiated pulse optimization.}
    \label{fig:auto-diff}
\end{figure}

In this work, we used auto-differentiation to optimize the smooth pulse ansatz, and the reported fidelities and population trajectories were obtained using the optimized pulse shapes in Qutip. In Fig.~\ref{fig:auto-diff}, we show an illustration of the auto-differentiated pulse optimization workflow.

\bibliography{main}
\end{document}